\documentclass[letterpaper,twocolumn,10pt]{article}
\usepackage{usenix2019,epsfig,endnotes}
\usepackage{fullpage}

\usepackage{hyperref}
\usepackage{url}        
\usepackage{float}
\usepackage{graphicx}
\usepackage{amsmath}
\usepackage{color}
\usepackage{times}
\usepackage{enumitem}
\usepackage[font={scriptsize,it}]{subfig}
\usepackage{verbatim}
\usepackage{xspace}
\usepackage{xfrac}
\usepackage{algorithm, algpseudocode}
\usepackage[normalem]{ulem}
\usepackage{amsfonts}
\usepackage{lastpage}
\usepackage{boxedminipage2e}
\usepackage{authblk}

\usepackage[export]{adjustbox}

\usepackage[small,compact]{titlesec}
\usepackage[utf8]{inputenc}
\usepackage{cleveref}
\usepackage{listings}
\usepackage{microtype} 
\usepackage{balance}   
\usepackage{datetime}
\usepackage{morefloats}

\usepackage[font={footnotesize, it}]{caption}
\usepackage{multirow}

\usepackage{tikz}

\input{defs}

\DeclareMathVersion{mathchartertext}

\begin{document}

\date{}

\title{Tiramisu: Fast and General Network Verification}

\author{
	{\rm Anubhavnidhi Abhashkumar$^*$, Aaron Gember-Jacobson$^{\dag}$, Aditya Akella$^*$}\\
	University of Wisconsin - Madison$^*$, Colgate University$^{\dag}$ \\
}

\maketitle
\thispagestyle{empty}
{\bf Abstract:} Today's distributed network control planes support
multiple routing protocols, filtering mechanisms, and route selection
policies. These protocols operate at different layers, e.g. BGP
operates at the EGP layer, OSPF at the IGP layer, and VLANs at layer
2. The behavior of a network's control plane depends on how these
protocols interact with each other. This makes network configurations
highly complex and error-prone. State-of-the-art control plane
verifiers are either too slow, or do not model certain features of the
network. In this paper, we propose a new multilayer hedge graph
abstraction, \Name, that supports fast verification of the control
plane. \Name uses a combination of graph traversal algorithms and ILPs
(Integer Linear Programs) to check different network policies. We
use \Name to verify policies of various real-world and synthetic
configurations. Our experiments show that Tiramisu can verify any
policy in $<$ 0.08 s in small networks ($\sim$35 devices) and $<$ 0.12
s in large networks ($\sim$160 devices), and it is 10-600X faster than
state-of-the-art without losing generality.

\section{Introduction and Background}

Many networks, including university, enterprise, ISP and data center
networks, employ complex distributed control planes running atop rich
underlying network structures. The control planes run a variety of
routing protocols, such as RIP, OSPF, and BGP, which are configured in
intricate ways to exchange routing information both within and across
protocols. In some cases, the protocols are assisted by {\em other}
protocols (e.g., iBGP assisting in distributing BGP information
throughout a network). Networks also employ techniques to virtualize
the control plane, such as virtual routing and forwarding (VRF), as
well as techniques to virtualize multiple network links into different
broadcast domains (VLANs).

Bugs can easily creep into such networks, e.g.,
through errors in the detailed configurations that many of the
protocols need. Often, bugs may not be immediately apparent, and the
network may be running ``just fine'' until a failure causes a latent
bug to be triggered. When bugs do manifest, a variety of catastrophic
outcomes can arise: the network may suffer from blackholes; services
with restricted access may be rendered wide open; critical
applications can no longer be isolated from other
services/applications, and so on.

{\bf Verification tools and their trade-offs.} A variety of tools
attempt to verify if networks are violating, or could potentially
violate, policies of the above kind. Data plane
verifiers~\cite{anteater,hsa,veriflow} analyze the current forwarding
tables and check for blackholes, loops, or broken path
isolation. Unfortunately, these tools don't have the means to analyze
if the network's new data plane that materialized upon reacting to a
failure, can satisfy relevant policies or not.

To overcome this issue, a variety of {\em control plane analyzers}
were developed~\cite{batfish,minesweeper,bagpipe,era,arc}. These
proactively analyze the network against various environments, e.g.,
failures or external advertisements. While a significant step forward
in network verification, control plane tools today make trade-offs
between performance and generality.

On the one hand are ``graph-based'' tools such as ARC~\cite{arc}. ARC
encodes all paths that may manifest in a network under various
failures into a series of weighted digraphs. This abstraction enables
analyzing the network under many potential environments at once by
running very fast polynomial time graph algorithms; e.g., checking if
two hosts are always blocked amounts to checking if they are in
different graph connected components. Unfortunately, ARC ignores many
network design constructs, including modeling the intricacies of BGP
and iBGP, and the existence of VLANs, and VRFs.

On the other hand are ``SMT-based'' tools such as
Bagpipe~\cite{bagpipe} and Minesweeper~\cite{minesweeper}. These tools
create a detailed model of the control plane by symbolically encoding
routing information exchange, route selection logic, and the
environment (e.g., failures) using logical constraints/variables. By
asking for a satisfying assignment for a SMT formula that encodes
the network and the property of interest, they can identify a concrete
environment that leads to property violation. These tools offer much
better coverage of control plane protocols than graph-based ones, but
their verification performance is very poor, especially when exploring
failures (\secref{sec:eval}), despite many internal SMT-specific
optimizations.

{\bf Decoupling encoding and properties.} We ask if it is possible to
design a verification tool that marries the speed of graphs with the
generality of an SMT-based encoding.

We start by noting that today the trade-off between performance and
generality in tools, arises from a coupling between the control
plane encoding used in the tools and how properties are verified. In
graph-based tools, the weighted graph control plane model requires
graph algorithms to verify properties. In SMT-based tools,
the detailed constraint-based control plane encoding requires a
general constraint solver to be used for all properties.

Our framework, \Name, decouples the encoding from the property: it
uses a sufficiently rich encoding for the network that models various
control plane features and network design constructs. But then, it
permits the use of custom algorithms that offer the best performance
to verify a property of interest.

{\bf Richer graphs.}  Our framework starts with graphs as the basis
for modeling networks, because graph-based control plane analysis has
been shown to be fast~\cite{arc}. We then embellish the graph model
with richer graph constructs, such as hierarchical layering, the
notion of hedges~\cite{hedgegraph} (edges that fail together), and
rich, multi-attribute edge and node labels. Our resulting graph can
model all aspects we observe used in real world networks and
configurations.

Given the graph structure, we then develop a suite of custom
techniques that help verify various properties of interest.
  
{\bf Avoiding path enumeration.} First, we note that some properties
of interest do not require the computation of the actual path that the
network would induce on a certain failure; they care mostly about
whether paths exist or not.  An example is whether two hosts are
always blocked from each other. For such properties, we develop two
techniques that avoid path enumeration altogether: a modified depth
first search graph traversal algorithm, and a simple integer linear
program (ILP) formulation that computes graph cuts. Importantly, graph
traversal runs in polynomial time. And, for a given property, the simplified property-specific ILP
can be solved much faster than a general SMT-encoding in SMT-based
tools. This is because the former 
explores symbolically only the variables that are relevant to the
property being verified, whereas the SMT solver for the general encoding searches through a
much larger search space.

{\bf Domain-specific path computation.} For the remaining properties
that require computation of paths, we run a modified path vector
protocol atop our richer graph abstraction. Here, we leverage
foundational work by Griffin {\em et. al} which showed that various
routing protocols can be modeled as instances of the stable paths
problem~\cite{griffin2002stable} (this insight was used in
Minesweeper), and that a ``simple path vector protocol''
(SPVP)~\cite{griffin2002stable} emulates the 
computation of a solution to this problem. In our version
of SPVP, each node consumes the multi-dimensional attributes of incoming edges and
neighboring vertices, and uses simple arithmetic operations that
encode existing protocols' logic to select among multiple paths
available at the moment. Importantly, in simple networks (e.g., those
that use a single routing protocol) our protocol naturally devolves
to being similar to distance vector protocol, which runs in polynomial
time (being based on the Bellman-Ford algorithm). For more general
networks, we empirically find that our protocol can quickly compute
the paths that are relevant to verifying specific properties. The
performance is faster than SMT-based tools, because our approach
essentially uses a highly domain-specific approach to finding paths,
compared to a SMT-based general search strategy; for instance, even
for simple networks, a SMT-based strategy would invoke the solver to
find a satisfying assignment.

{\bf Prototype and evaluation.} We implemented \Name in Java (7K LOC)
and evaluated it with many real data center and university
networks, as well as networks from the topology zoo~\cite{zoo}. We find that
\Name's rich multi-layered graphs can be computed from configurations
in a few $\mu$s per traffic class. Using \Name's custom algorithms,
various properties can be checked for complex networks in 3-80ms per
traffic class. Compared to Minesweeper that uses a general encoding with an SMT solver, \Name offers speed of up to 600X for
reachability policy verification and 10-50X for bounded-length and
path preference policies (both under failures). \Name's algorithmic
approach renders it substantially faster even when verifying
properties under no failures. Finally, \Name scales well, providing
verification results in $\sim$100ms per traffic class for networks
with $\sim$160 routers.

\section{Motivation} \label{sec:motivation}

\setlength{\textfloatsep}{0.3cm}
\setlength{\floatsep}{0.3cm}

Given the significant performance benefits of graph
abstractions~\cite{arc}, we use graphs as the basis to encode control
plane computation in \Name. However, the abstraction we use is
significantly different from ARC~\cite{arc}. In what
follows, we provide an overview of ARC's graph
based approach for verification. We then identify its key drawbacks
which motivate \Name's graph design.

\subsection{ARC}
ARC (Abstract representation for control plane~\cite{arc}) is a
state-of-the-art control plane verifier. It models a network's control
plane using a collection of directed graphs. There is one directed
graph per traffic class which models the forwarding behavior of
packets belonging to that traffic class. In ARC, nodes represent
routing processes, and directed edges represent possible flow of
traffic enabled by exchange of route advertisements between routing
processes. Using a {\em single attribute} edge-weight, ARC can model
OSPF costs and AS\_path length. Finally, ARC verifies policies by
checking some simple graph characteristics. \tabref{tab:pol} lists
some of these policies and related graph properties. By leveraging
graphs, ARC offers order-of-magnitude better performance~\cite{arc}
compared to state-of-the-art~\cite{batfish,minesweeper}.

However, ARC's drawback is that it is limited in its network design
coverage. It does not model layer 3 protocols such as iBGP and VRF. It
does not model any layer 2 protocols including VLANs. It also does not
model BGP protocol attributes such as local preference, communities
etc. We use examples to show how these limitations can affect ARC's
correctness, and what we do in \Name to overcome them.

\setlength{\textfloatsep}{0.1cm}
\setlength{\floatsep}{0.1cm}

\begin{table}
\begin{centering}
  \footnotesize
  \setlength\tabcolsep{3pt}
\begin{tabular}[b]{p{0.4\columnwidth} | p{0.5\columnwidth}}
\hline
Policy class & Graph characteristics \\
\hline
\bpolicyBlock $src$ and $dst$ are always block & $src$ and $dst$ are in separate components\\
\bpolicyWay All paths from $src$ to $dst$ traverses a waypoint & after removing the waypoint, $src$ and $dst$ are in separate components\\
\bpolicyFail $src$ can reach $dst$ when there are $<$ K link failures & min-cut of $src-dst$ graph is $>=$ K \\
\hline
\end{tabular}
\end{centering}
\caption{Policies as graph characteristics}
\label{tab:pol}
\vspace*{-2em}
\end{table}
\setlength{\textfloatsep}{0.4cm}
\setlength{\floatsep}{0.4cm}

\begin{figure}
	\includegraphics[scale=0.45]{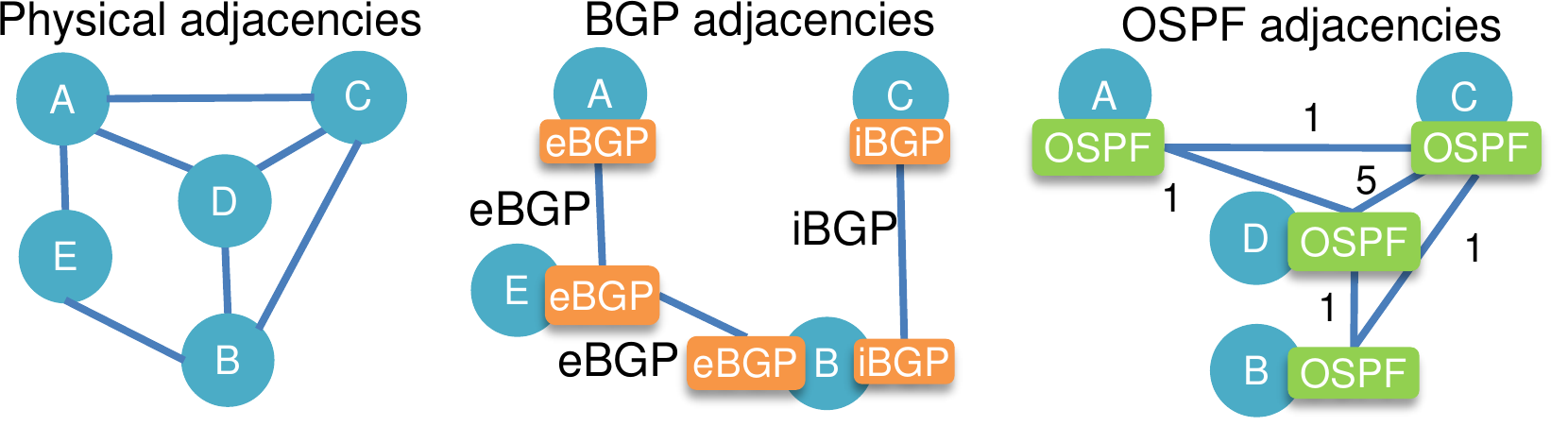}
	\compactcaption{Multilayer dependency}
	\label{fig:ibgp_ospf}
\end{figure}

\begin{figure}
	\centering
	\includegraphics[scale=0.45]{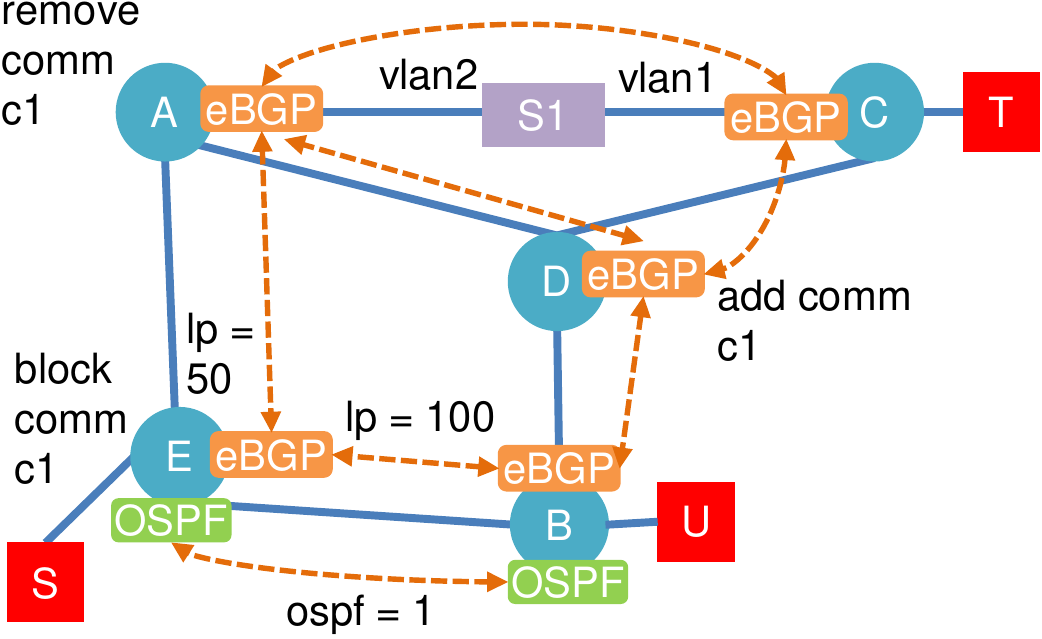}
	\compactcaption{Multiple network features}
	\label{fig:protocols}
\end{figure}

\textbf{Cross layer dependency.} Consider the network
in~\figref{fig:ibgp_ospf}. The network consists of 5 routers
($A-E$). Here, routers $A$ and $B$ are $eBGP$ peers of router $E$,
and, router $C$ is an $iBGP$ peer of router $B$. All routers except
$E$ belong to the same $OSPF$ process, with each link except $C-D$
having a link cost of 1. Link $C-D$ has cost 5. In this scenario, $C$
learns a route to $E$ through its $iBGP$ neighbor $B$, with $B$ as
its next hop. $C$ forwards all traffic destined to $E$, to
$B$. $iBGP$ peers forward traffic at the $IGP$ layer. Hence $C$ uses
its $OSPF$ process to find the path to reach $B$. The path computed
varies for different failure scenarios, which affects reachability of
traffic. Depending on the path used under a given failure, the
operator's policy for traffic between C and E may be violated:\\ i)
Under no failures, $OSPF$ prefers path $C \rightarrow B$ (OSPF cost
1), and then traffic flows from $B \rightarrow E$;\\ ii) When the link
$B \rightarrow C$ fails, $OSPF$ prefers a different path $C
\rightarrow A \rightarrow D \rightarrow B$ (cost 3). Crucially,
traffic at $A$ gets directly rerouted to $E$, because $A$ is the
$eBGP$ peer of $E$; \\ iii) When links $A \rightarrow C$ and $B
\rightarrow C$ fail, $OSPF$ prefers path $C \rightarrow D \rightarrow
B$ (cost 6). Here traffic gets dropped at $D$, because $D$ did
not learn a route for $E$, and $C$ is disconnected from
$A$.  Note that, had router $A$ or $B$ redistributed it's $BGP$ routes
to $OSPF$, then $D$ would have learned a route to $E$, avoiding the
blackhole situation (iii) above.

Such dependencies between $iBGP$ and $OSPF$ cannot be
modeled in ARC. ARC cannot analyze these scenarios and compute the
actual paths used, and because of it, it cannot be used to verify any
policies in this network.

ARC's lack of modeling of cross-layer dependencies impacts its
applicability in other network scenarios too. Consider the network
in~\figref{fig:protocols}. This network has 5 routers
($A-E$), a switch $S1$, and three hosts $S-U$. All
routers run BGP, and routers $B$ and $E$ also run OSPF (with cost
1). Switch $S1$ connects to router $C$ and $A$ on VLAN 1 and 2,
respectively. $D$ adds community ``c1'' to its advertisements, $A$
removes ``c1'', and $E$ blocks all advertisements with community
``c1''. Finally, to prefer routes learned from $B$ over $A$, $E$
assigns local preference values $100$ and $50$ to $B$ and $A$,
respectively.

Although routers $A$ and $C$ are connected to the same switch $S1$,
they belong to different VLANs. Hence, traffic cannot flow through
switch $S1$. By default, ARC assumes layer 2 connectivity. Hence
according to ARC routers $A$ and $C$ are reachable and traffic can
flow between them.

The overall theme is that protocols ``depend'' on each
other. E.g. iBGP depends on OSPF, BGP and OSPF depend on VLANs
etc. These protocols also operate at different ``layers''. BGP
operates at the EGP (Exterior Gateway Protocol) layer, OSPF at the IGP
(Interior Gateway protocol) layer, and VLANs at Layer 2. A graph
abstraction needs to encode layers and cross-layer dependencies. Thus,
\Name introduces a new multilayer graph abstraction, where traffic
flow in the higher layers may depend on the traffic flow in the lower
layers. \figsref{fig:correlated}{fig:mulgraph} show the multilayer
graphs of the aforementioned networks, which we explain in detail
in~\secref{sec:multigraph}

\textbf{Impact of BGP attributes.} In~\figref{fig:protocols}, the
path taken from $E$ to $C$ depends on communities and local preference. There are two paths
from $E$ to $C$: i) $E \rightarrow B \rightarrow D \rightarrow C$,
and, ii) $E \rightarrow A \rightarrow D \rightarrow C$. Because of
local preference, path (i) is preferred over path (ii). However, $E$
blocks all advertisements with community ``c1''. Since router $D$ adds
this community, all advertisements through $D$ will have community
``c1''. Hence the advertisement for path (i) is blocked. Although the
advertisement for path (ii) also comes through $D$, router $A$ removes
community ``c1'' and router $E$ does not see that
community. Therefore, there is only one path between $E$ and $C$:  $E \rightarrow A \rightarrow D \rightarrow C$. ARC, on
the other hand, characterizes path (i) as valid.

Additionally, ARC cannot model other BGP attributes like local
preference, MED etc. To support all BGP attributes as well as metrics
of other protocols (OSPF cost, AD), \Name uses multiple edge and node
attributes. We elaborate in~\secref{sec:multigraph}.

\textbf{Physical link dependency.} Consider the network
in~\figref{fig:protocols} without link $A-E$. Here, according to ARC,
traffic from $S$ to $U$ can flow through 2 paths $S \rightarrow
E_{ospf} \rightarrow B_{ospf} \rightarrow U$ and $S \rightarrow
E_{bgp} \rightarrow B_{bgp} \rightarrow U$. To evaluate reachability
under failure, ARC calculates {\em min-cut} of this graph as 2, and
concludes that it can withstand arbitrary 1 link failures. However,
this is incorrect because \edge{E_{ospf}}{B_{ospf}} and
\edge{E_{bgp}}{B_{bgp}} belong to the same underlying physical link
$E-B$ whose failure causes disconnection.

In graphs, the notion of hedges~\cite{hedgegraph} can be used to model
such dependency among edges. A hedge is a set of edges that fail
together. A graph can have multiple hedges, each with more than one
edge in it, and a single edge can belong to multiple hedges. In \Name,
to support edges that fail together, we label sets of ``related''
edges as hedges. Overall, \Name converts the control plane into a {\em
  series of multilayer multi-attribute hedge graphs} one per
source-destination pair.

\section{\Name Graph Abstraction} \label{sec:multigraph}

\begin{figure}
	\includegraphics[scale=0.45]{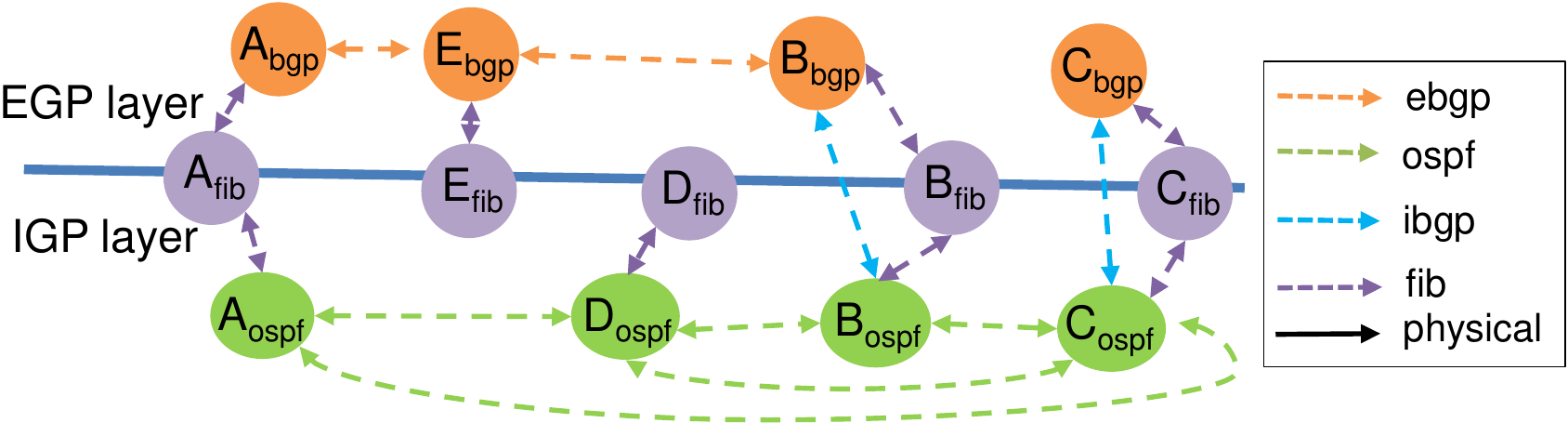}
	\compactcaption{Correlated Edges}
	\label{fig:correlated}
\end{figure}

\begin{figure}
	\includegraphics[scale=0.45]{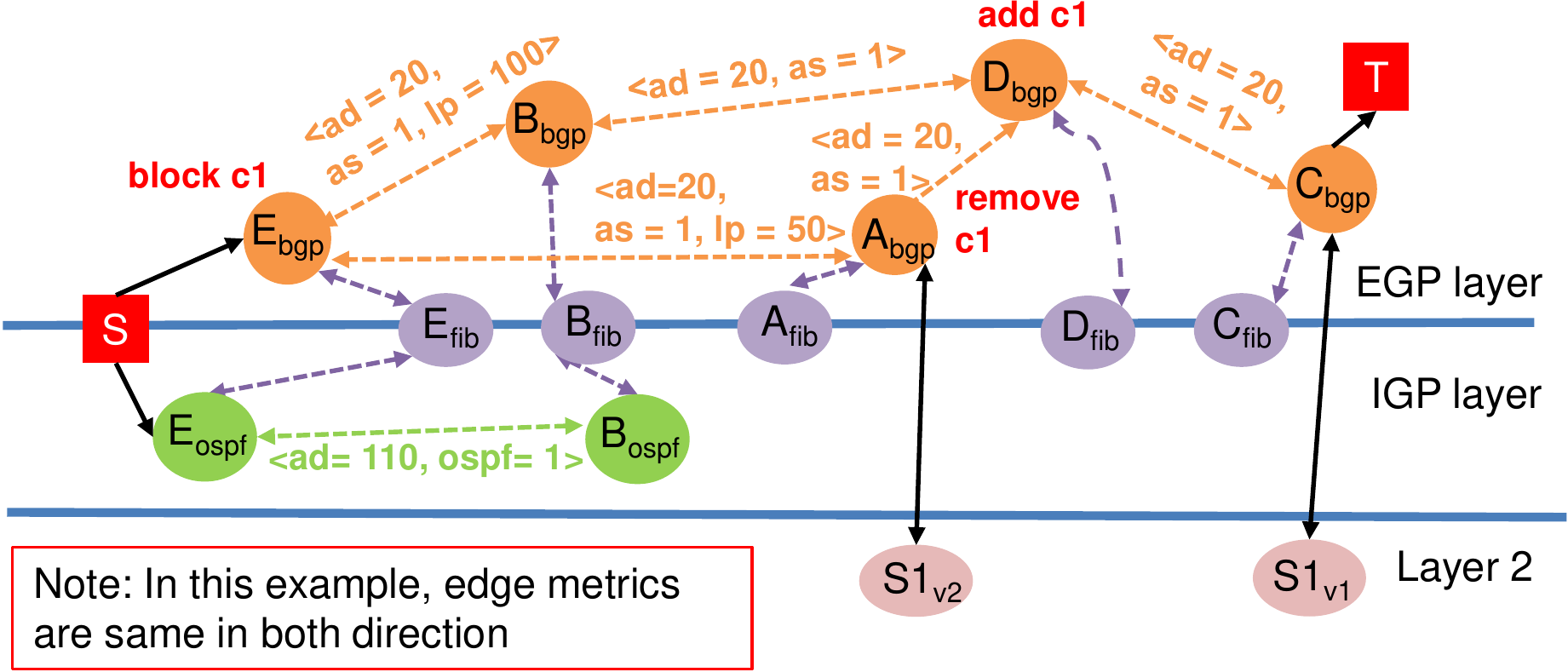}
	\compactcaption{Multilayer Hedge Graph for traffic class $S-T$}
	\label{fig:mulgraph}
\end{figure}

We discuss the components of the \Name graph abstraction.

\textbf{Nodes.} In \Name, nodes are created for both switches and
routers. For switches, nodes are created per device, per VLAN. E.g,
in~\figref{fig:mulgraph}, $S2$ has two nodes $S2_{v1}$ and $S2_{v2}$
for VLANs $1$ and $2$. For routers, nodes are created per device, per
routing process. E.g, in~\figref{fig:mulgraph}, router $B$ has two
nodes $B_{bgp}$ and $B_{ospf}$ for it's BGP and OSPF routing
processes. If the router supports VRF (Virtual Routing and
Forwarding), then the nodes are replicated per device, per virtual
routing process (similar to VLAN). Note that by default, each routing
process has a default\_vrf, and nodes like $B_{bgp}$ represent the
default\_vrf of B's BGP process. We identify routing processes, VLANs,
and VRFs from device configurations. We also create, per router, a
node representing the router's forwarding information base
(FIB). Finally, we create two special nodes representing the $src$ and
$dst$ of a traffic class.

\textbf{Edges.} \Name categorizes it's edges into multiple labels,
depending on the edge's end nodes. These labels include: FIB (f), eBGP (b), IGP (o), static (s), redistribute (r),
iBGP (i), and physical (p). Some of these labels are for inter-device
edges and some for intra-device edges. We show examples of these
labels in~\figref{fig:mulgraph}, and explain them below.

\textbf{Inter device edges.} An inter-device edge exists between
different devices that have a physical link connecting them. There are
three cases: (i) if both end nodes are switches and belong to the same
VLAN, connect them with an edge of label ``p'', (ii) if one node is a
switch and other a router, connect them with edge of label ``p'',
e.g. \edge{S1_{v1}}{A_{bgp}}. (iii) if both nodes are routers, they
belong to the same routing process (IGP or eBGP) and the same VRF,
then connect them with edge label ``o'' or ``b'', respectively,
e.g. \edge{D_{bgp}}{A_{bgp}}.

\textbf{Intra device edges.} An intra-device edge exists between
nodes that belong to the same device. There are three cases: (i) if
the router redistributes routes from process $X$ to $Y$, add edge of
label ``r'' from $Y$ to $X$, e.g. \edge{B_{bgp}}{B_{ospf}} , (ii) if
the router is an iBGP peer, add edge of label ``i'' from the router's
BGP process to the underlying IGP process,
e.g. \edge{B_{bgp}}{B_{ospf}} of~\figref{fig:correlated}, or (iii)
every node corresponding to a routing process on a given router is
connected (both ways) to the router's FIB node; e.g., $B_{ospf}$ is
connected to $B_{fib}$ (\figref{fig:correlated}).


Each traffic class has a $srcRouter$ and $dstRouter$ that originate
the $src$ and $dst$ IP addresses. \Name connects the $src$ node to all
routing processes of $srcRouter$. Finally, it connects all routing
processes of $dstRouter$ with $dst$ node.

\textbf{Edge costs.} \Name supports multiple routing
protocols. Each protocol uses a different set of \emph{metrics} to
express link and path costs/preferences. E.g. OSPF uses link costs,
and, BGP uses AS-path length, local preference (lp), Multi-Exit
Descriptor (MED) etc. Additionally, an administrative distance ($AD$),
allows operators to choose routes from different protocols. Hence a
single edge weight cannot model all protocols. \Name replaces these
edge weights with a \emph{vector of metrics}. The values of these
metrics are inferred from device configurations. Depending on the edge
label, certain metrics will be set as null, e.g. OSPF cost is null for
``b'' edges.

\textbf{Hedges.} All edges that represent the same physical link
belong to the same hedge. E.g. in~\figref{fig:mulgraph},
\edge{B_{bgp}}{E_{bgp}} and \edge{B_{ospf}}{E_{ospf}} belong to the
same physical link $B \rightarrow E$, and thus the same hedge.

\textbf{Traffic class specific constructs.}: Some aspects of the
network are specific to traffic classes, e.g. ACLs, filters, static
routes, etc. Thus, \Name first creates a single {\em base graph}
representing the above features/protocols for the entire
network. From that, it then creates {\em traffic class-specific} graphs.

ACLs prevent specific traffic classes from entering/leaving a router,
and filters block advertising specific prefixes. Thus, for a given
traffic class, \Name logically removes an edge of the traffic class graph if
there is (i) an ACL that blocks this traffic class on the interface
associated with this edge, or, (ii) a filter that blocks this traffic
class's destination prefix on the routing process associated with the
start node of this edge.

Static routes are also traffic class specific.  \Name adds an edges of
label ``s'' from the node representing the source of the static route
process to all nodes associated with the next hop router, for the
relevant traffic class.

\textbf{Node communities.} Communities are also a traffic class
specific construct. In \Name, each node representing a $BGP$ process
has three sets of communities: ``ac'' - communities added by the node,
``rc'' communities removed by the node, and ``mc'' communities
matched/acted upon by the node. The action associated with each ``mc''
is to either block traffic or change edge metrics like local
preference and MED.

\subsection{Prohibited paths}\label{subsec:prohibit}

All the algorithms we present to verify various properties rely on
examining properties of source-destination paths in the above graph
abstraction. In designing these algorithms care must be taken to avoid
certain paths that {\em cannot} materialize in a real network under
{\em any} failure. This mainly arises due to constraints on
interactions between routing protocols and route
redistributions/static routes, and due to communities. We show how we
reason about which routes cannot be taken due to the former reason. We
handle communities later.

For example, as shown in~\secref{sec:motivation}, traffic crossing
from an EGP to the IGP layer gets dropped if the intermediate nodes at
the IGP layer do not learn/have a route to $dst$. 

Intuitively, a path can materialize in the network if there is
relevant forwarding information available toward the destination in
the RIB of some routing process running on each router on the path,
and thus in the router's FIB; this observation forms the basis for the
seminal work on static reachability analysis of IP
networks~\cite{xie2005static}. A path cannot materialize if it
includes a router with no forwarding information for the
destination. To keep track of such disallowed paths for graph
traversal or when computing paths taken, \Name uses a ``tainting''
strategy. In a preprocessing step, taints track how forwarding
information may flow through a network's RIBs and may populate FIBs.
Taint on a node implies that a node {\em may} know of a route to a
destination; lack of a taint implies the node {\em will not} know of
any route. 

For each traffic class, taints propagate in the corresponding \Name
graph starting at the destination $dst$; the routing process connected
to the $dst$ is tainted, as are all other peer processes of the same
routing protocol instance on other routers. Taints flow across
redistribution and static route edges, and spread throughout the
vertices corresponding to a single IGP. Taints also propagate from one
iBGP peer to another (because iBGP peers learn of routes from each
other).

Specifically, to identify nodes in the IGP layer that have
a route to $dst$, \Name first finds the node at the EGP layer that
redistributes its route to IGP, and {\em taint}s this node.\footnote{Note,
that edges in \Name represent the flow of traffic and not of
advertisements. Hence, redistribution of EGP routes into IGP is
represented by an ``r'' edge {\em from} the IGP node {\em to} the EGP
node.} Next, \Name marks all nodes in the IGP process's layer as {\em
  taint}ed, implying that they may know of a route to $dst$ in their
RIB due to redistribution from the EGP (recall: IGPs flood
learned routes, to routers in the same IGP process). On the
other hand, if there is no redistribution from EGP into IGP, the
corresponding IGP node is marked as {\em untaint}ed, because it {\em
  will not} not have a route to the destination in its RIB (but a
route may exist in the router's FIB because the EGP process computed a
route); all other nodes in the same IGP process are also {\em
  untaint}ed. E.g. in~\figref{fig:correlated}, all OSPF nodes are untainted.

The above processes is applied to all routing layers and EGP/IGP
crossings, of which there may be many in a network. In the end, we
have a tainted graph, with a subset of vertices carrying taints and
others without taint; the latter vertices will not have a forwarding
entry to the destination.


Given a tainted graph, we determine which paths are disallowed.
Intuitively, an untainted node must reside on the same router as a
tainted node for a path to the destination to be found in the router's
FIB. Thus, a potentially valid path has $\le 2$ consecutive
untainted nodes (e.g., the two IGP nodes). Any path with $>2$
consecutive untainted nodes is an invalid or \emph{prohibited
  path}. E.g. in~\figref{fig:correlated}, $C_{ospf} \rightarrow
D_{ospf} \rightarrow A_{ospf}$ is a subpath of prohibited paths.

Although our example considered iBGP, IGP and BGP interactions, taints
can also be used to understand how static routes can shape the flow of
routing information, and consequently impact paths.
\vspace*{-0.4em}
\begin{theorem}\label{theorem:proh}
The above approach correctly identifies all prohibited path.
\end{theorem}
\vspace*{-0.4em}
We prove this by contradiction as shown in~\appref{subsec:proh}.

A potentially valid path may still not materialize in the network
due  communities. We handle communities specially
in our  algorithms in subsequent sections.



\section{Avoiding Path Enumeration}\label{sec:pathset}

For many properties of interest, verifying them does not actually
require computation or enumeration of actual paths that may arise in
the network; rather, we simply need to know of the existence of paths
(possibly of a certain kind). Knowing of the existence of paths is far
simpler than actually computing paths  under a given failure.

Thus, we design two sets of algorithms, one that traverses a graph
along all potential paths, and another that uses ILPs to reason about
the number of paths and high-level path properties (maximum
path length). We present these algorithms in the context of the
properties they aid in the verification of. All algorithms
use taints to take invalid paths out of consideration, and handle
communities specially as mentioned above.


\subsection{Tiramisu Depth First Search}\label{subsec:tdfs}

$DFS$ (Depth First Search) is a graph traversal algorithm that
identifies all nodes that are reachable from a given $src$. A node
remains unvisited after $DFS$ iff there exists no path from $src$ to
that node. The DFS algorithm can be naturally used to verify certain
reachability policies, such as, \policyBlock (``always blocked''):
\policyBlock is true iff under every failure scenario, there exists
             {\em no path} between $src$ and $dst$ node. However
             standard DFS does not avoid \emph{prohibited path}
             constraints, nor does it account for how BGP communities
             may shape the existence of
             paths. E.g. in~\figref{fig:correlated}, paths with
             subpath $C_{bgp} \rightarrow C_{ospf} \rightarrow
             B_{ospf} \rightarrow D_{ospf}$ are prohibited. But $DFS$
             identifies them as valid. 

\setlength{\textfloatsep}{0.1cm}
\setlength{\floatsep}{0.1cm}
\begin{algorithm}[t]
\small
\caption{Tiramisu Depth First Search}
\label{alg:tdfs}
\begin{algorithmic}[1]
\Statex{\textbf{Input:}}
\INDSTATE{$G$ is the graph}
\INDSTATE{$src$ is the root node for depth first search}
\Procedure{$initialize(G, src)$}{}
  \State{Set all nodes of $G$ as unvisited}
  \State{$TDFS$($G$, $src$, 0)}
\EndProcedure

\Statex{\textbf{Input:}}
\INDSTATE{$u$ is current node being traversed}
\INDSTATE{$numUntaint$ tracks \# consecutive untainted nodes}
\Procedure{$TDFS$(G, u, $numUntaint$)}{}
  \State{Set node $u$ as visited}
  \ForEach {$e \in G.outgoingEdge[u]$}
    \State{v $\gets$ end node of edge e}
    \State{Ignore v, if already visited}
    \If{$v$ is untainted}
      \State{increment $numUntaint$ by 1}
      \State{Ignore v, if $numUntaint$ is 3}
    \Else
      \State{$numUntaint \gets$ 0}
    \EndIf
    \State{$TDFS$(G, v, $numUntaint$)}
  \EndFor
\EndProcedure

\end{algorithmic}

\end{algorithm}

We first assume no communities are in use and propose a modified DFS
algorithm (\algoref{alg:tdfs}) that avoids prohibited paths, \tdfs (Tiramisu DFS). We
then account for communities by invoking \tdfs on carefully selected
subgraphs.

Like DFS, \tdfs explores all unvisited outbound neighbors of a node
(line 6 to 8). However, it uses an additional variable $numUntaint$ to
avoid traversing prohibited paths.

In this algorithm, $numUntaint$ represents the number of consecutive
untainted nodes seen by \tdfs; \tdfs simply avoids all paths with more
than 2 consecutive untainted nodes (line 11). E.g., in \figref{fig:mulgraph},
 when \tdfs reaches node $D_{ospf}$, $numUntaint$ becomes 3
and \tdfs stops exploring this path. 
\tdfs's complexity is the same as
standard DFS.

\tdfs algorithm does not handle communities. Nodes may add communities
to path advertisements that others filter on. When such filtering
happens, the corresponding paths cannot be taken by network traffic.
To support communities, \Name uses another simple algorithm, \commtdfs
(\algoref{pseudo:comm}). \commtdfs makes a constant number of calls to
\tdfs verify \policyBlock. Thus, its overall complexity is still
polynomial-time.

As shown in~\secref{sec:motivation}, nodes can add, remove or block on
communities. In presence of such nodes, the order in which these nodes
are traversed in a path decides if $dst$ is reachable.

In the \commtdfs algorithm, \Name first checks if $src$ and $dst$ are
unreachable according to \tdfs (line 5, 6). If they are reachable,
then \commtdfs checks if all paths that connect $src$ to $dst$ has i)
a node (say X) that blocks on a community in an advertisement for
$dst$ (line 8 to 10), ii) followed by a node (say Y) that adds that
community to an advertisement for $dst$ (line 11 to 13), iii) and no
node between them X and Y that removes that community (line 14 to
16). (Recall here that advertisements flow in the opposite direction
of data traffic.) If all these conditions are satisfied, then $src$
and $dst$ are unreachable. If any of these conditions are violated,
the nodes are reachable.
In~\figref{fig:mulgraph}, $E_{BGP}
\rightarrow A_{BGP} \rightarrow D_{BGP}$ violates condition (iii).

\setlength{\textfloatsep}{0.1cm}
\setlength{\floatsep}{0.1cm}
\begin{algorithm}[t]
\small
\caption{Always blocked with communities}
\label{pseudo:comm}
\begin{algorithmic}[1]
\Statex{\textbf{Input:}}
\INDSTATE{$G$ is the graph}
\INDSTATE{$s$ and $d$ are source and destination nodes}
\INDSTATE{It uses \tdfs to answer multiple (un)reachability queries}
\Procedure{$\commtdfs(G, s, d)$}{}
  \State{$cA \gets$ nodes that add community}
  \State{$cR \gets$ nodes that remove community}
  \State{$cB \gets$ nodes that block on community}
  \If{$s$ and $d$ are unreachable by \tdfs}
  	\State{\Return true, since nodes are already unreachable}
  \Else
  	\State{remove nodes $\in$ $cB$}
  	\If{$s$ and $d$ are reachable by \tdfs}
  		\State{\Return false, since there is path between $s$ and $d$ that is not blocked by community}
  	\EndIf
  	\State{add back all nodes, remove nodes $\in$ $cA$}
  	\If{nodes $\in$ $cB$ and $d$ are reachable by \tdfs}
  		\State{\Return false, since blocking nodes can receive advertisement without community}
  	\EndIf
  	\State{add back all nodes, remove nodes $\in$ $cR$}
  	\If{nodes $\in$ $cB$ and their respective nodes $\in$ $cA$ are unreachable by \tdfs}
  		\State{\Return false, since communities are always removed before reaching blocking nodes}
  	\EndIf
  	\State{Return true, if all above conditions fail}
  \EndIf
\EndProcedure

\end{algorithmic}
\end{algorithm}

This algorithm can also verify \policyWay (``always waypointing''). After removing the waypoint, if $src$ can
reach $dst$, then there is a path that can reach $dst$ without traversing the waypoint.

\subsection{\Name Hedge Min-cut}\label{subsec:min-cut}

Another policy that does not require path enumeration is 
\policyFail (reachable $<$ K failures): here, operators want to verify
that the $src$ can reach $dst$ as long as there are $< K$ link
failures. For this, ARC~\cite{arc} computes the 
min-cut for the $src$-$dst$ graph, i.e., the minimum number of edges
whose removal can disconnect the graph. If min-cut is $\ge K$, then
\policyFail is satisfied.

\Name uses a multilayer hedge graph abstraction. Unfortunately,
finding min-cuts in hedge graphs is a known NP-hard
problem~\cite{hedgegraph}. Thus, to verify \policyFail, we propose a
new Integer Linear Program (ILP) to find hedge graph min-cut, while
also accounting for communities and prohibited paths.

For this property, \Name uses integer variables, similar to
Minesweeper~\cite{minesweeper}'s SMT encoding.  However, \Name's
property-specific ILP encoding is simpler than Minesweeper as it only
adds integer and boolean variables relevant to hedge min-cut and
avoids variables and constraints associated with route selection,
which significantly reduces variables and constraints associated with
route advertisements. Minesweeper's detailed encoding exercises all of
these variables in an attempt to compute the actual path, which
renders it slow even for a small number of failures ($K=1$).


Until now, the edges in our graphs represented the flow of traffic
from $src$ to $dst$. For ease of understanding, in specifying the ILP,
we reverse the edges to represent the flow of advertisement from $dst$ to
$src$. For brevity, we explain the constraints at a
high-level, leaving precise definitions to
~\appref{subsec:min-cut-appendix}. Equation numbers below refer
to equations in~\appref{subsec:min-cut-appendix}.

The objective of the ILP is to minimize the number of physical link
failures required to disconnect the $src$ from $dst$. Note that a
single physical edge's failure can cut multiple \Name graph edges. 
All such edges
belonging to hedge $i$ will share the same $F_i$ variable. Our objective then is: 

\begin{equation} \label{eq:obj}
\textbf{Objective:} \tab  \texttt{minimize} \sum_{i \in  phyEdges} F_i
\end{equation}

\textbf{Advertisement Constraints.} We first discuss the
constraints added to represent reachability of advertisements. The
base constraints state that the $dst$ originates the advertisement (\eqnref{eq:dst}). To
disconnect $dst$ from $src$, the advertisement must not reach the
$src$ (\eqnref{eq:src}). For other nodes, the constraint is that an advertisement reaches a
node if it gets propogated on any of its incoming edge
(\eqnref{eq:reach_cons}).





Next, we discuss constraints on propogation of advertisements. An
advertisment can propogate through an edge $e$, if it reaches the
start node of the edge, the physical edge does not fail, the
advertisement does not carry a community that is blocked on this node,
and if edge $e$ carries atleast one advertisement that does not create
a prohibited path. This is represented as shown in
\eqnref{eq:flow_cons}.


\textbf{Communities.} The base constraints state that
each node that adds a community forwards that community and each
node that removes that community does not forward it further
(\eqnref{eq:add_comm} and \eqnref{eq:rem_comm}).



For other nodes, we add the constraint that node $n$ forwards a
community $c$ iff any of its inbound neighbors forwards that
community to node $n$ (\eqnref{eq:oth_comm}).

Finally, we add the constraint that an edge $e$ carries a blocked
community iff the start node of $e$ forwards a community that
is blocked by the end node of edge $e$ (\eqnref{eq:block_comm}).


\textbf{Prohibited Path Constraints.} An edge $e$, that only
propogates advertisements that create prohibited paths, is an edge
that satisfies the condition that start and end nodes of $e$ are
untainted ($uEdges$) (\eqnref{eq:proh}). Also, the start node receives
advertisement only from untainted neighbors and not from any other
neighbor. Such edges always create subpaths with three consecutive
untainted nodes (\secref{subsec:prohibit}) (\eqnref{eq:proh}).





\subsection{\Name Longest path}\label{subsec:bound}

Similar to cut-based properties, there are others that don't need path
enumeration and can instead rely on custom-crafted ILPs. These
properties compute high-level attributes of paths, such as bounds on
length. An example is the always bounded length policy
(\policyBound). For a given $K$, \policyBound is true if under every
failure scenario, traffic from $src$ to $dst$ never traverses a path
longer than $K$ hops. Enumerating all possible paths and finding their
length is infeasible. However, this policy can be verified efficiently
by viewing it as a variation of the {\em longest path problem}: for a
given $K$, \policyBound is true if the longest path between $src$ and
$dst$ in the graph is $\le K$.

\textbf{ILP.} Finding the longest path between two nodes in a
hedge graph is also NP hard~\cite{karger1997approximating}. To verify \policyBound, we propose
another ILP whose objective is to maximize the number of inter device
edges ($dEdges$) traversed by an advertisement ($A_i$). Note again
that the path traversed by the advertisement is the opposite of
traffic flow. But both have same path length. Notably, this ILP uses
even fewer constraints and variables compared
to~\secref{subsec:min-cut}, and thus can run even faster relative to
Minesweeper.

\begin{equation} \label{eq:obj2}
\textbf{Objective:} \tab  \texttt{maximize} \sum_{i \in dEdges} A_i
\end{equation}

Again, we elide detailed constraints to the~\appref{subsec:bound-appendix}.

\textbf{Single Path Constraints.} We add constraints to ensure
that only one path gets advertised, that the $dst$ sends the
advertisement, and that the $src$ receives it
(\eqnref{eq:dst2} and \eqnref{eq:src2}).



For other nodes, we add the flow conservation property, i.e. sum of incoming flows is equal to outgoing flows (\eqnref{eq:flowconser}).


\textbf{Advertisement Constraints.} 
Finally, we add constraints on propogation of advertisements. Here, an advertisement can be blocked on edge $e$ if carries a blocked community or satisfies the prohibited path constraints (\eqnref{eq:extracons}).


\section{Path specific policies}\label{sec:path}

As opposed to the properties in the previous section, the remaining
properties that one may wish to verify require knowing the exact
path taken in the network. Consider the policy that encodes a
preference order among paths, say, $prefPath$ $=$ $P1 >> P2 >>
P3$. This states that $P1$ should be taken by default; when path $P1$
fails, path $P2$ (if available) should be taken; and if $P1$ and $P2$
both fail, then $P3$, if available, should be taken. A path, say $P1$,
can become unavailable for multiple reasons: (a) a router
configuration along $P1$ was updated to withdraw or filter the route to the
$dst$; (b) a link along $P1$ failed. In all cases, we need to reason about what
alternate paths materialize, and whether the materialized path is
indeed the path that was meant to be taken. Simple graph traversal is thus
insufficient to verify this property. We need the actual paths 
under various failure scenarios.

Griffin \emph{et al}~\cite{griffin2002stable} observed that control
plane protocols essentially attempt to solve the stable paths
problem. Based on this, to compute paths,
Minesweeper~\cite{minesweeper} models protocol interactions
(advertisement generation/processing, best path selection, etc) as an
instance of the stable paths problem. Furthermore, Griffin \emph{et
  al}~\cite{griffin2002stable} proposed the Simple Path Vector
Protocol (SPVP) for obtaining a set of paths that form a solution to
the stable path problem.

Inspired by these two studies, we propose \pvp (Tiramisu PVP), a protocol that can
compute actual network paths taken, while avoiding prohibited
paths. \pvp extends SPVP with rich messages and per-node
computation. 

Whereas Minesweeper encodes the router's path selection actions based
on SMT constraints, \pvp models them using a 
protocol. Thus, \Name emulates the {\em execution} of SPVP, whereas
Minesweeper attempts to find the SPVP {\em output}.

To find the path traversed by a specific traffic class, \Name runs
\pvp on its {\em traffic class-specific} graph.

\subsection{\Name Path Vector Protocol}

\textbf{Advertisements.}  Similar to SPVP, for each node $u$, and for
each of its peer $v$, \pvp uses \rib{u}{v} variables to keep
track of the most recent route advertisement received from $v$. Each
advertisement consists of (i) a path from the advertisement sender to the advertisement source ($dst$ of the traffic
class), and (ii) the multi-metric path cost to reach 
$dst$. Also, \ribs{u} represents the current path (best advertisement and best path cost) to reach $dst$ from $u$.

\textbf{Route Selection.}
In Minesweeper, the values of \ribingen were computed by applying import and export filters on these advertisements. In \pvp, these are replaced by an $updateCost$ function. This function (i) rejects prohibited paths (\secref{subsec:prohibit}), (ii) rejects paths blocked by communities  (\secref{subsec:tdfs}), and, (iii) computes the multi-atrribute path cost to reach $dst$ through $v$. Each node has a $choices$ function ($\oplus$) that provides a partial order over all \rib{u}{*} paths. $u$ then updates its \ribs{u} value to the best choice/most preferred \rib{u}{*}. 

\ribingen also keeps track of the list of communities carried by the advertisements, and the $updateCost$ function updates the community lists.

The $updateCost$ and $choices$ functions are closely modeled based on the route import, route export and route selection constraints of Minesweeper and SPVP. These functions are either inferred from configuration or based on standard conventions. E.g. $choices$ could be based on BGP selection policy (prefer highest $localpref$, shortest $ASPathlength$ etc), and, $updateCost$ specifies how to calculate the cost of \rib{u}{v} as a function of \best{v} and edge weight \ew{u}{v} (add $ospf costs$, set $localpref$ to \ew{u}{v}$.lp$ etc).

\textbf{Algorithm.}
Our final \pvp protocol is shown in~\algoref{alg:tpvp}. In the initial state, \pvp sets the \ribgen value of all nodes except $dst$ to null (line 2). \ribs{dst} is set to $\epsilon$, since it originates the advertisement (line 3). There are three main steps in each iteration. First, for each node $u$, \pvp computes all its \ribingen values based on the advertisement sent by its neighbors (line 6 to 8). If the current \ribs{u}'s value is different from the best advertisement, then \pvp updates \ribs{u} and sends a message to all peers of node $u$ (line 9 to 11). This message contains the updated \ribs{u}. 

\pvp terminates when none of the nodes receive any messages about the new \ribingen values (line 12, 13). This termination condition represents the converged state of the newtork. Similar to SPVP, if network reaches a converged state, then \pvp will find the best path to $dst$ for all nodes.

\begin{theorem}\label{theorem:tpvp}
Assuming the network control plane converges, \pvp always finds the exact path
taken in the network under any given failure scenario.
\end{theorem}

We leverage correctness of Minesweeper and SPVP to establish the
correctness of \pvp as shown in~\appref{subsec:tpvp_proof}.

We return to the correctness of \tdfs.

\begin{theorem}\label{theorem:tdfs}
\tdfs traverse all real network paths that could materialize under
some failure scenario, and does not traverse any path that cannot materialize on any
failure scenario.
\end{theorem}

We leverage the correctness of \thrref{theorem:proh} and \thrref{theorem:tpvp} to prove 
correctness of \thrref{theorem:tdfs} as shown in~\appref{subsec:tdfs_proof}. 


\setlength{\textfloatsep}{0.1cm}
\setlength{\floatsep}{0.1cm}
\begin{algorithm}[t]
\small
\caption{Tiramisu Path Vector Protocol}
\label{alg:tpvp}

\begin{algorithmic}[1]
\Statex{\textbf{Input:}}
\INDSTATE{$G$ is the graph}
\INDSTATE{$dst$ is the destination node}
\Procedure{$TPVP(G, dst)$}{}
  \State{set rib(i) values of all nodes $i$ except $dst$ to null}
  \State{set rib(dst) to $\epsilon$, since $dst$ originates the advertisement}
  \While{true}
    \ForEach {$u \in G.nodes$}
      \ForEach {$v \in peers(u)$}
        \State{$rib-in(u\Leftarrow v)$ = $updateCost(rib(v), \ew{u}{v})$}
        \State{compute $best(u)$ using $choices([rib-in(u\Leftarrow *)])$}
        \If{$rib(u)$ is different from $best(u)$}
          \State{$rib(u)$ = $best(u)$}
          \State{send messages to all peers of node $u$ about change in $rib(u)$}
        \EndIf
      \EndFor
    \EndFor
    \If{no node receives any messages}
      \State{break, since network has converged}
    \EndIf
  \EndWhile
\EndProcedure

\end{algorithmic}

\end{algorithm}

\subsection{Tiramisu Yen's algorithm}\label{subsec:yen}

We return to determining how to verify the path preference property
introduced earlier in this section. We first observe that there are
similarities between analyzing path preference (\policyPref) and
finding the $k$ shortest paths in a graph (the ``$k$ shortest paths
problem''~\cite{yen}).
This is because, in $k$ shortest paths problem, the $k^{th}$ shortest
path is taken only when $k-1$ paths have failed. Enumeration for all
possible failures of all $k-1$ shorter paths is tedious. To handle
that, Yen~\cite{yen} introduced a new algorithm to find the $k$
shortest paths. The algorithm uses dynamic programming to avoid
enumerating failures/link removals. Next, we explain Yen's algorithm
in detail. Then, we propose \tyen (\algoref{alg:tyen}), a simple
extension to Yen's algorithm to verify \policyPref. We show how in a
simple fashion \tyen uses Yen's algorithm to verify \policyPref.

\textbf{Yen's algorithm.} Yen uses two lists: $listA$ (keeps track of
the shortest path seen so far) and $listB$ (keeps track of candidates
for the next shortest path). At the start, Yen finds the first
shortest path $P$ (line 2) from $src$ to $dst$ using any shortest path
algorithm (e.g. Dijkstra's).

Next, it takes every node $n$ in path $P$ (line 11, 12), and finds the
$rootPath$ and $spurPath$ of that node. The $rootPath$ of $n$ is the
subpath of $P$ from $src$ to node $n$ (line 13). The $spurPath$ is the
shortest path from node $n$ to $dst$ (line 18) that satisfies the
following two conditions: i) the path must not have any node from
$rootPath$ (line 17); ii) the path must not traverse any outgoing edge
$e$ from $n$ that is part of any of the previous $k-1$ shortest paths
having the same $rootPath$ (line 15, 16). E.g. if $A \rightarrow B
\rightarrow C \rightarrow D$ is the shortest path and $A \rightarrow
B$ is $rootPath$ at node $B$, then $B \rightarrow C \rightarrow E \rightarrow D$
$spurPath$ is invalid.

Yen combines the $rootPath$ and $spurPath$ to form a new path $nP$
(line 19). $nP$ is added to $listB$ if it doesn't already exist in
$listA$ or $listB$ (line 22). After traversing all nodes of path $P$,
Yen adds $P$ to $listA$ (line 24). Next, Yen picks the shortest path
from $listB$ and reruns the previous steps with this path as the new $P$. This continues till Yen finds
$k$ paths.

\begin{algorithm}[H]
\small
\caption{Tiramisu Yen}
\label{alg:tyen}
\begin{algorithmic}[1]
\Statex{\textbf{Input:}}
\INDSTATE{$G$ is the graph}
\INDSTATE{$src$, $dst$ are source and destination nodes}
\INDSTATE{$level$ is no. of path specified in path-preference policy}
\INDSTATE{$prefPath$, a map of preference level and path}
\Procedure{\tyen$(G, src, dst, level)$}{}
  \State{$P \gets$ path from src to dst returned by \pvp }
  \State{$P.level$ $\gets$ 1}
  \State{$P.eRemoved$ $\gets$ [], as best path requires no edge removal}
  \State{$listA$ $\gets$ [], tracks paths already considered as $P$}
  \State{$listB$ $\gets$ [], tracks paths not yet considered}
  \Do
    \State{$mostPref \gets$ most preferred path in $prefPath$ whose edges don't overlap with $P.eRemoved$}
    \If{$P$ $\neq$ $mostPref$}
      \State{\Return false, since path preference is violated}
    \EndIf
  	\For{i $\gets$ 0 to P.length - 1}
  		\State{$sNode \gets i^{th}$ node of P}
  		\State{$rootPath \gets$ subpath of P from $src$ to $sNode$}
  		\ForEach {$sp \in$ $listA$} \Comment{paths in $listA$}
  			\If{$sp$ has same $rootPath$ at $sNode$}
  			\State{remove outgoing edge of $sNode$ in $sp$, so that path $sp$ is not considered} \Comment{also remove other edges that share the same hedge}
  			\EndIf
  		\EndFor 
  		\State{remove all nodes and edges (hedge) of $rootPath$ except $sNode$ to avoid loops}
      \State{$spurPath$ $\gets$ path from $sNode$ to $dst$ returned by \pvp}
  		\State{$nP \gets$ $rootPath$ + $spurPath$}
  		\State{$nP.level \gets P.level + 1$}
      \State{$nP.eRemoved \gets$ all edges removed in this iteration}
		\State{add nP to end of $listB$ if np is valid and np $\notin$ [$listA$, $listB$]}
		\State{add back all nodes and edges to the graph}
  	\EndFor
  	\State{add $P$ to $listA$}
  	\State{P $\gets$ remove first path from $listB$}
  \doWhile{$P.level < K$}
  \State{\Return true, since loop didn't find preference violation}
\EndProcedure
\end{algorithmic}
\end{algorithm}

\textbf{Tiramisu Yen.}  The three failure modes highlighted at the
beginning of this section boil down to some specific edges being
removed from the \Name graph, causing the path in question to cease to
exist. Given this, we make a few modifications to Yen, resulting in an
algorithm we call \tyen, to work atop our multi-graph abstraction and
check for path preference; simply put, like Yen, we check if $P2$ is
the next path chosen after $P1$, and so on (line 8 to 10).

\tyen uses \pvp instead of Dijkstra to find the shortest path. In
addition, \tyen associates each path with a variable, $eRemoved$. This keeps track of edges that were removed to prefer this path (line 21). During each iteration of $P$, \tyen identifies the most prefered path in $pathPref$ that did not have an edge in $P.eRemoved$ (line 8). If this path varies from $P$, then preference is violated (line 9, 10).

\textbf{Revisiting: Reachability $<$ K with ACLs.} The ILP in
\secref{subsec:min-cut} for \policyFail is not accurate when data
plane ACLs are in use. ACLs do not influence route
advertisements. Hence, routers can advertise routes for traffic that
ends up being blocked by ACLs. Recall that during graph creation,
\Name removes edges that are blocked by ACLs
\secref{sec:multigraph}. This leads to incorrect mincut
computation as we show next:

Consider a $src$-$dst$ traffic class in a network. Say there are three
network paths $P1$, $P2$ and $P3$ in increasing order of cost that
$src$ learns of toward $dst$, and say $P2$ has a data plane ACL on
it. Suppose further than all three paths are edge disjoint. If a link
failure removes $P1$, the control plane would select path
$P2$ to forward to $dst$, but all packets from $src$ are dropped at
the ACL on $P2$. In this case, a single link failure (that took down
$P1$) is sufficient to disconnect $src$ and $dst$; that is, the true
mincut is 1. In contract, \Name would remove the offending ACL edge
from the graph abstraction, and since this preserves paths $P1$ and
$P3$, \Name would conclude that the mincut is 2, which is incorrect.

We address this issue as follows. Nodes can become unreachable when
failures either disconnect the graph or lead to a path with an
ACL. Thus, we compute two quantities: (1) $L$: How many minimum
failures cause the control plane to pick a path that first encounters
a blocking ACL? (2) $N$: How many minimum failures cause disconnection
in the \Name graph with the ACL edge removed (as originally proposed)?
The true min-cut value is $min(L,N)$?

Computing $N$ in (2) is straightforward. For computing $L$ in (1), we
first construct a graph without removing edges for ACLs. Then, we run
\tyen until we find the first path with a dropping ACL on it. Say this
was the $M^{th}$ shortest. Then, we use an ILP to compute the minimum
number of edge failures $L$ that will cause the previous $M-1$
shortest paths to fail. If $min(L, N) \ge K$ then \policyFail is
satisfied.


So far, we covered five properties to verify. But, using the
algorithms in this and the previous section as the basis, \Name can
verify a variety of other properties which Minesweeper can also
verify. We list these, along with how \Name verifies them, in~\appref{sec:otherPols}.

\section{Evaluation}\label{sec:eval}

\begin{table}[t]
	\begin{centering}
		\footnotesize
		\setlength\tabcolsep{3pt} 
		\begin{tabular}{c | c c c}
			\hline
			& \multicolumn{3}{c}{\% of Networks} \\
			Protocols/Modifiers & University & Datacenter & Topology Zoo \\ 
			\hline
			eBGP & 100\% & 100\% & 100\%\\
			iBGP & 100\% & 0\% & 100\%\\
			OSPF & 100\% & 97\% & 100\%\\
			Static routes & 100\% & 100\% & 0\%\\
			ACLs & 100\% & 100\% & 0\%\\
			Route Filters & 100\% & 97\% & 100\%\\
			Local Prefs & 50\% & 0\% & 100\%\\
			MPLS+VRF & 100\% & 0\% & 0\%\\
			VLAN & 100\% & 0\% & 0\%\\
			Community & 100\% & 100\% & 100\%\\
			\hline
		\end{tabular}
	\end{centering}
	\caption{Configuration constructs used in networks}
	\label{tab:conf}
\end{table}

\begin{figure}
	\includegraphics[scale=0.25]{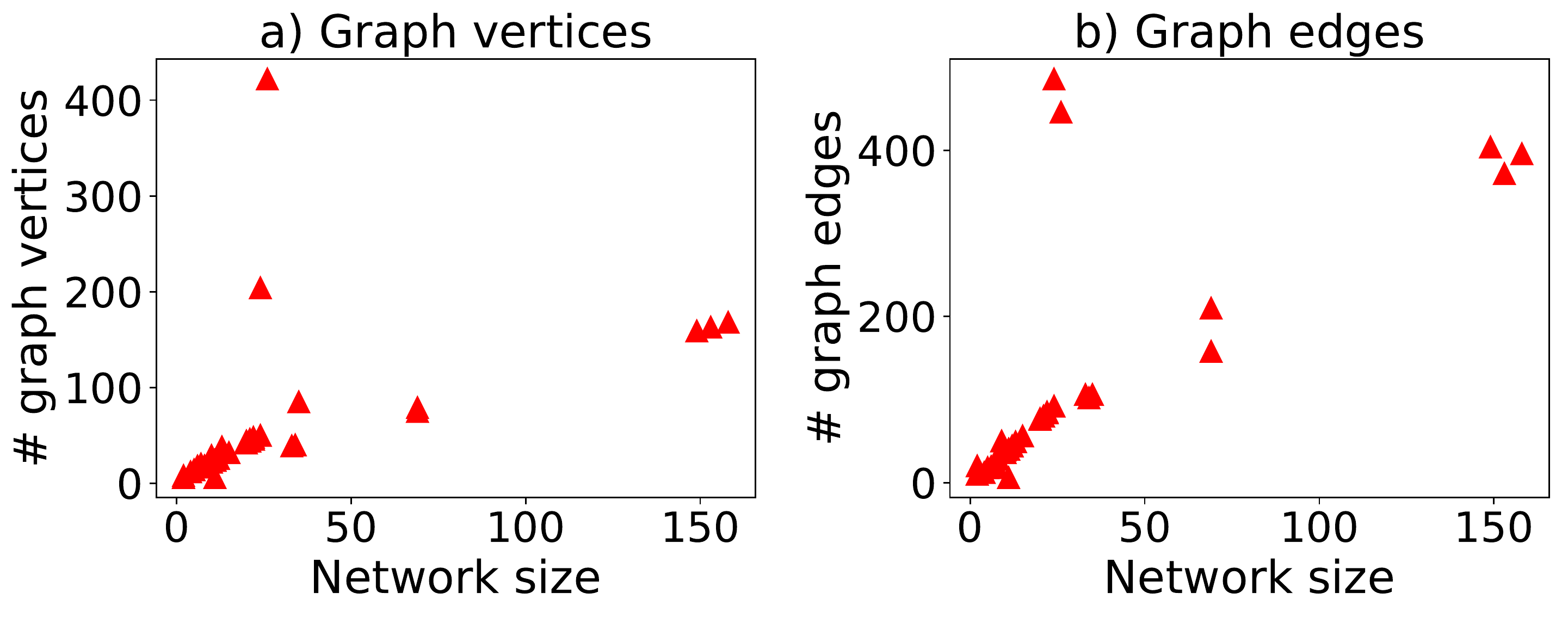}
	\vspace*{-0.4em}
	\caption{Size of multilayer graphs of all networks}
	\vspace*{-0.6em}
	\label{fig:graph}
\end{figure}

\begin{figure}
	\includegraphics[scale=0.25]{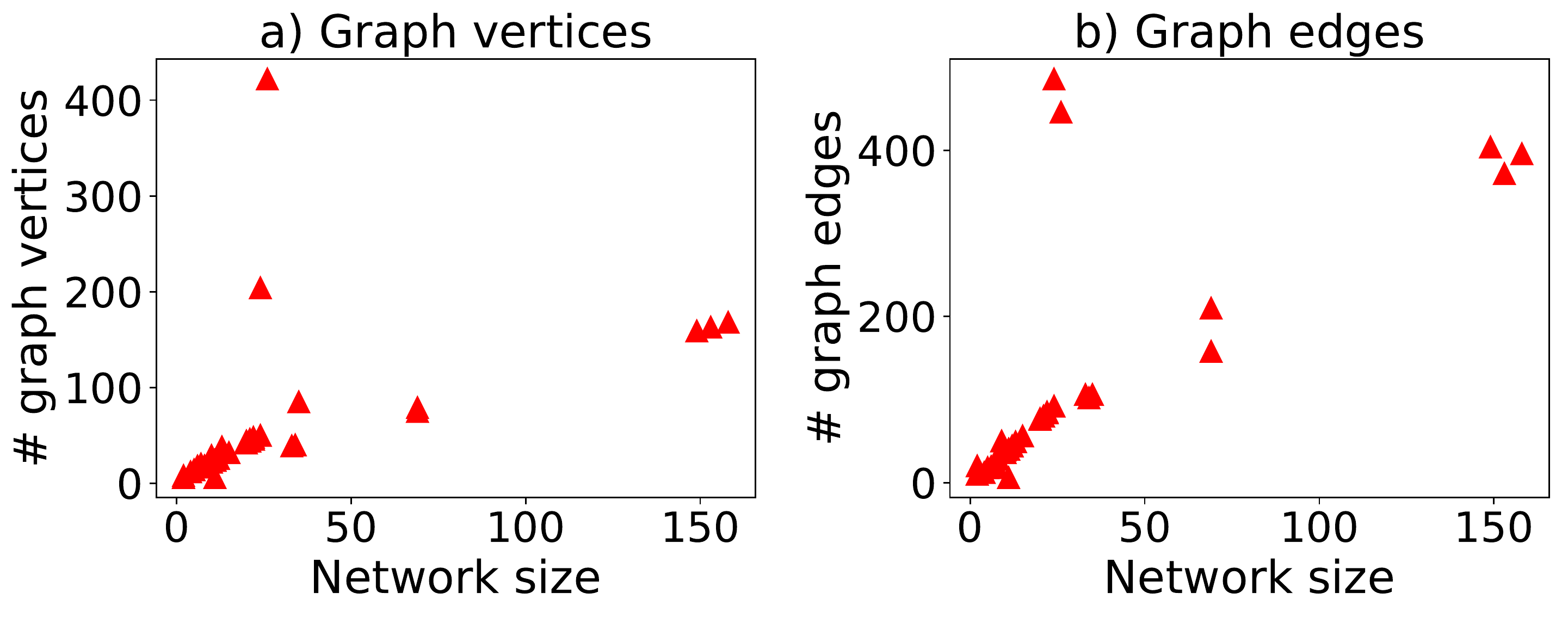}
	\vspace*{-0.4em}
	\caption{Size of multilayer graphs of all networks}
	\vspace*{-0.6em}
	\label{fig:graph}
\end{figure}

\begin{figure}
	\centering
  \includegraphics[scale=0.4]{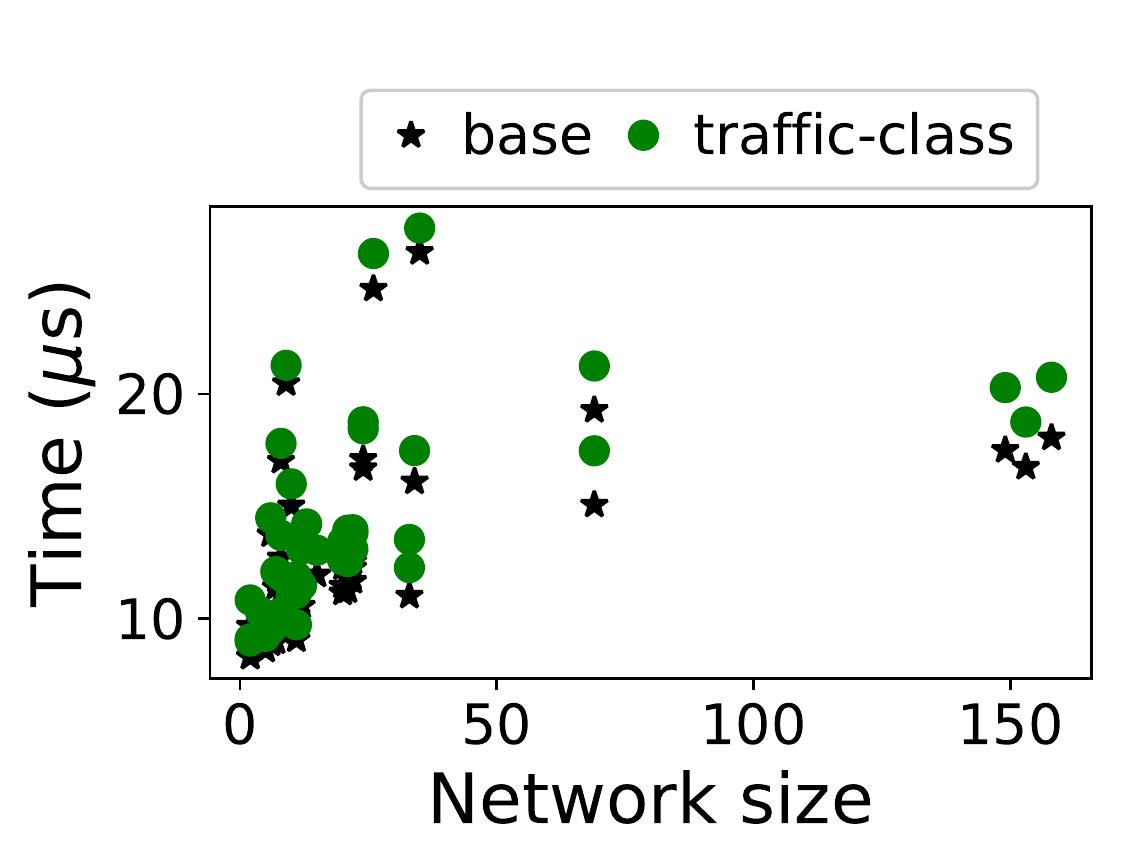}
\vspace*{-0.5em}
\caption{Graph generation time (all networks)}
\label{fig:gen}
\end{figure}

\begin{figure}
	\centering
  \includegraphics[scale=0.4]{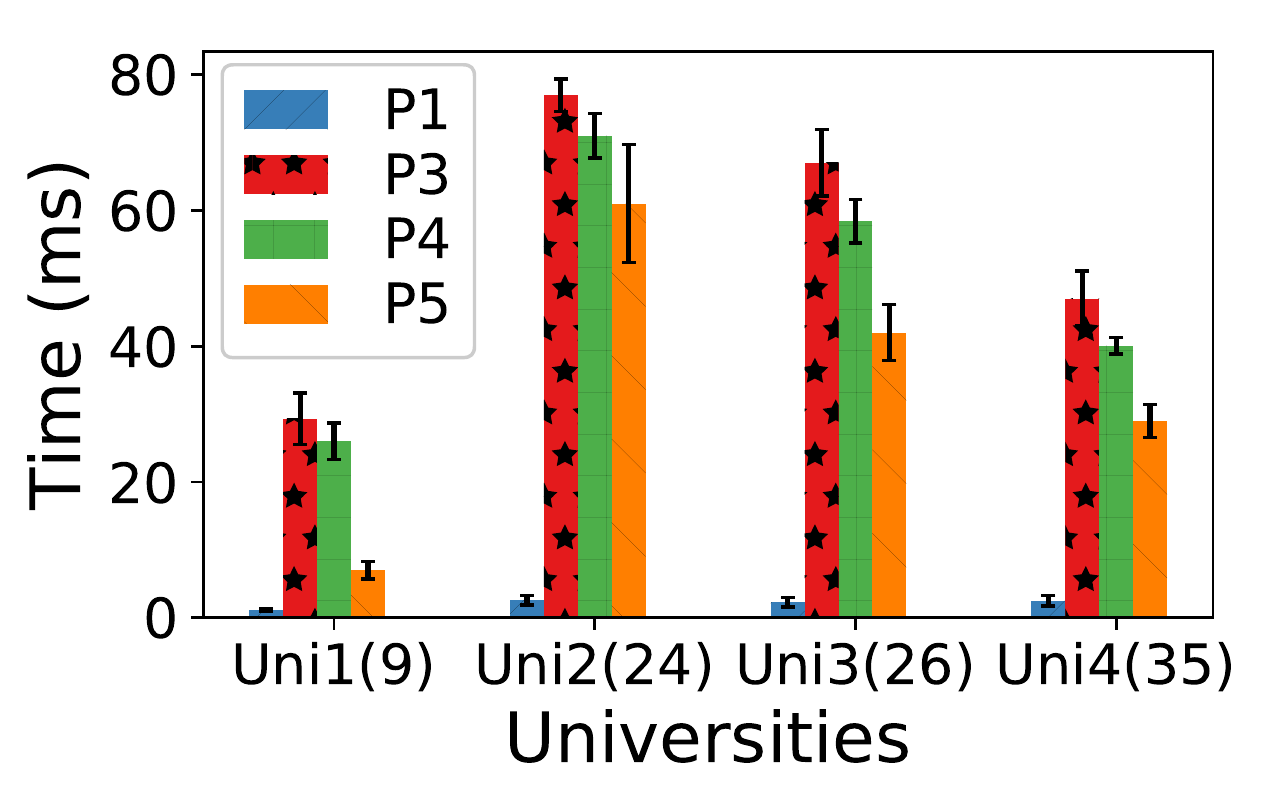}
\vspace*{-0.5em}
\caption{Verify policies on university configs}
\label{fig:univs}
\end{figure}

\begin{figure*}
	\includegraphics[width=1.8\columnwidth]{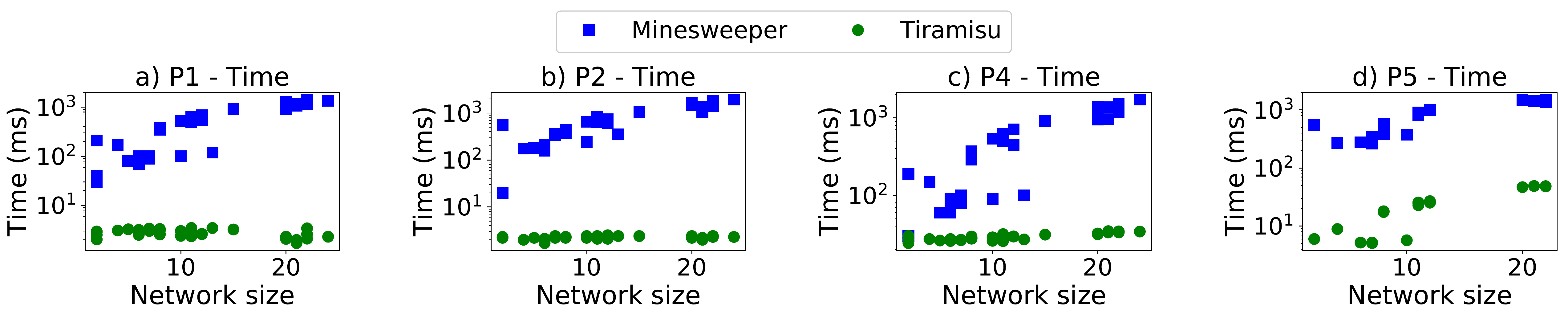}
	\compactcaption{Performance under all failures: \Name vs Minesweeper (datacenter networks)}
	\label{fig:minesweeper}
\end{figure*}

\begin{figure*}
	\includegraphics[width=1.8\columnwidth]{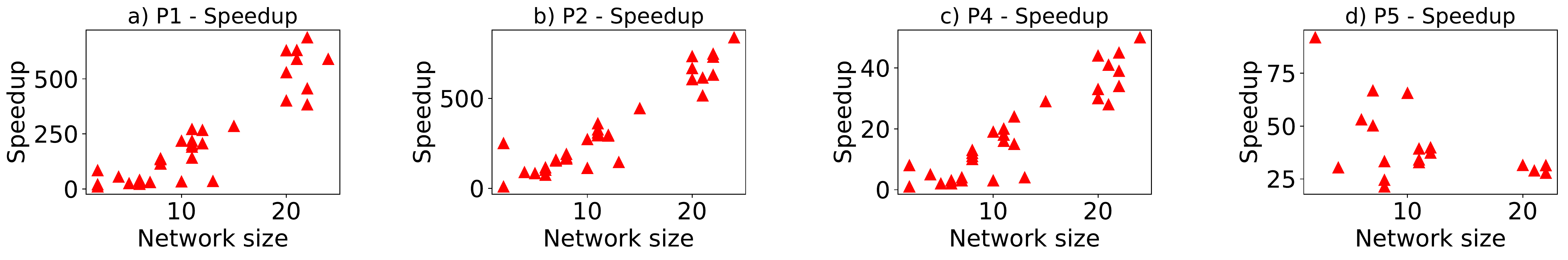}
	\compactcaption{Speedup under all failures: \Name vs Minesweeper (datacenter networks)}
	\label{fig:failspeed}
\end{figure*}

\begin{figure*}
	\includegraphics[width=1.8\columnwidth]{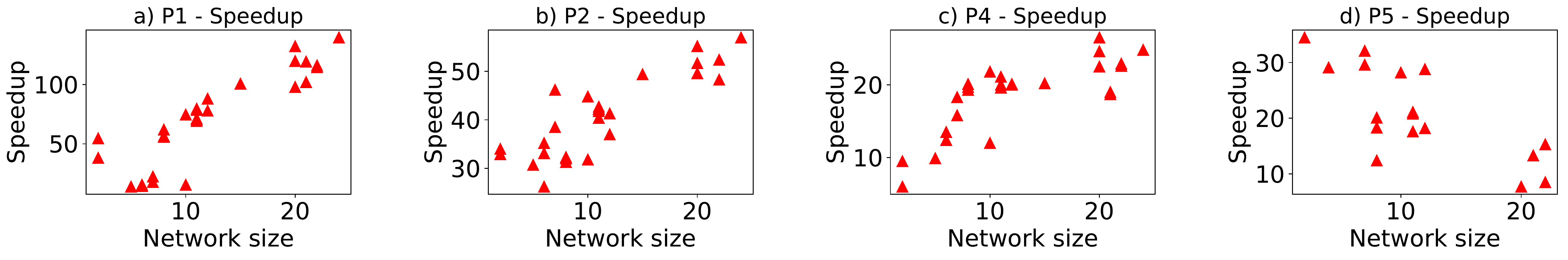}
	\compactcaption{Speedup under no failures: \Name vs Minesweeper (datacenter networks)}
	\vspace*{-0.6em}
	\label{fig:nofail}
\end{figure*}

\begin{figure}
	\centering
	\includegraphics[scale=0.35]{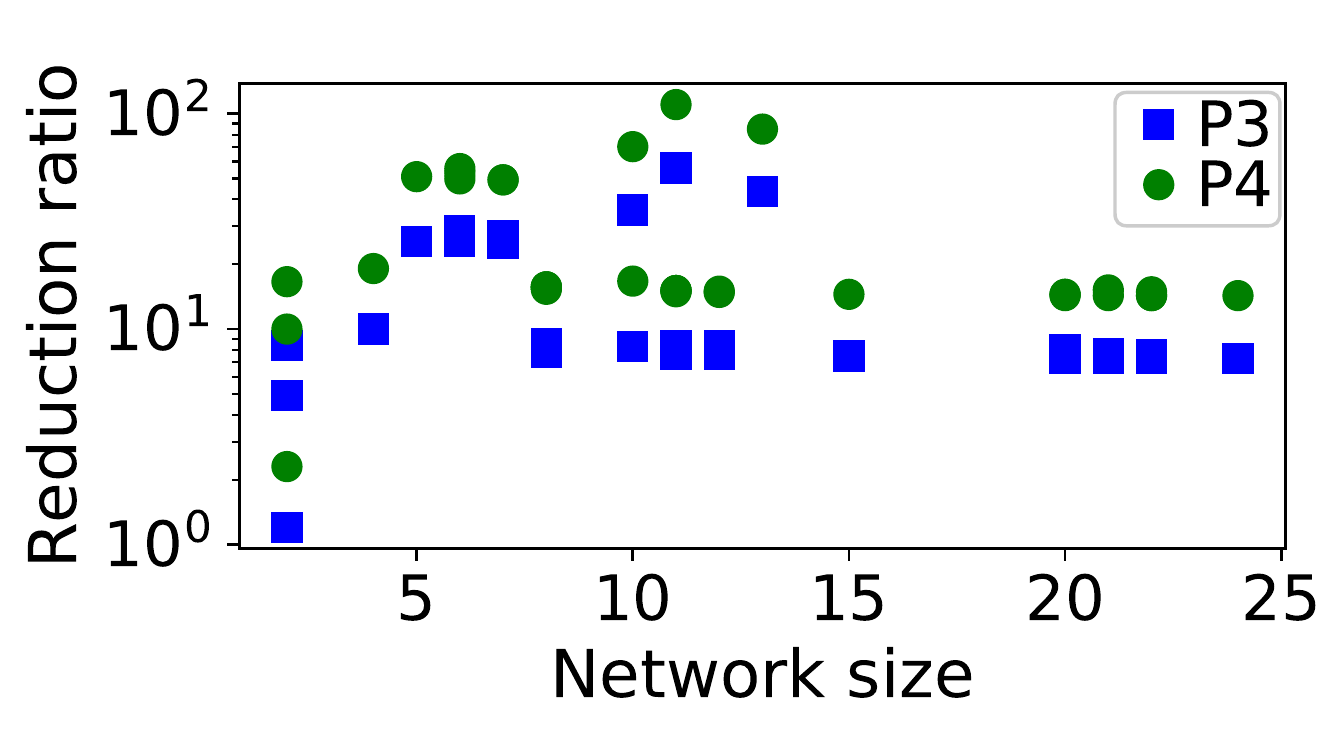}
	\compactcaption{Variable reduction ratio - \Name vs Minesweeper}
	\label{fig:vars}
\end{figure}

\begin{figure}
	\centering
	\includegraphics[scale=0.4]{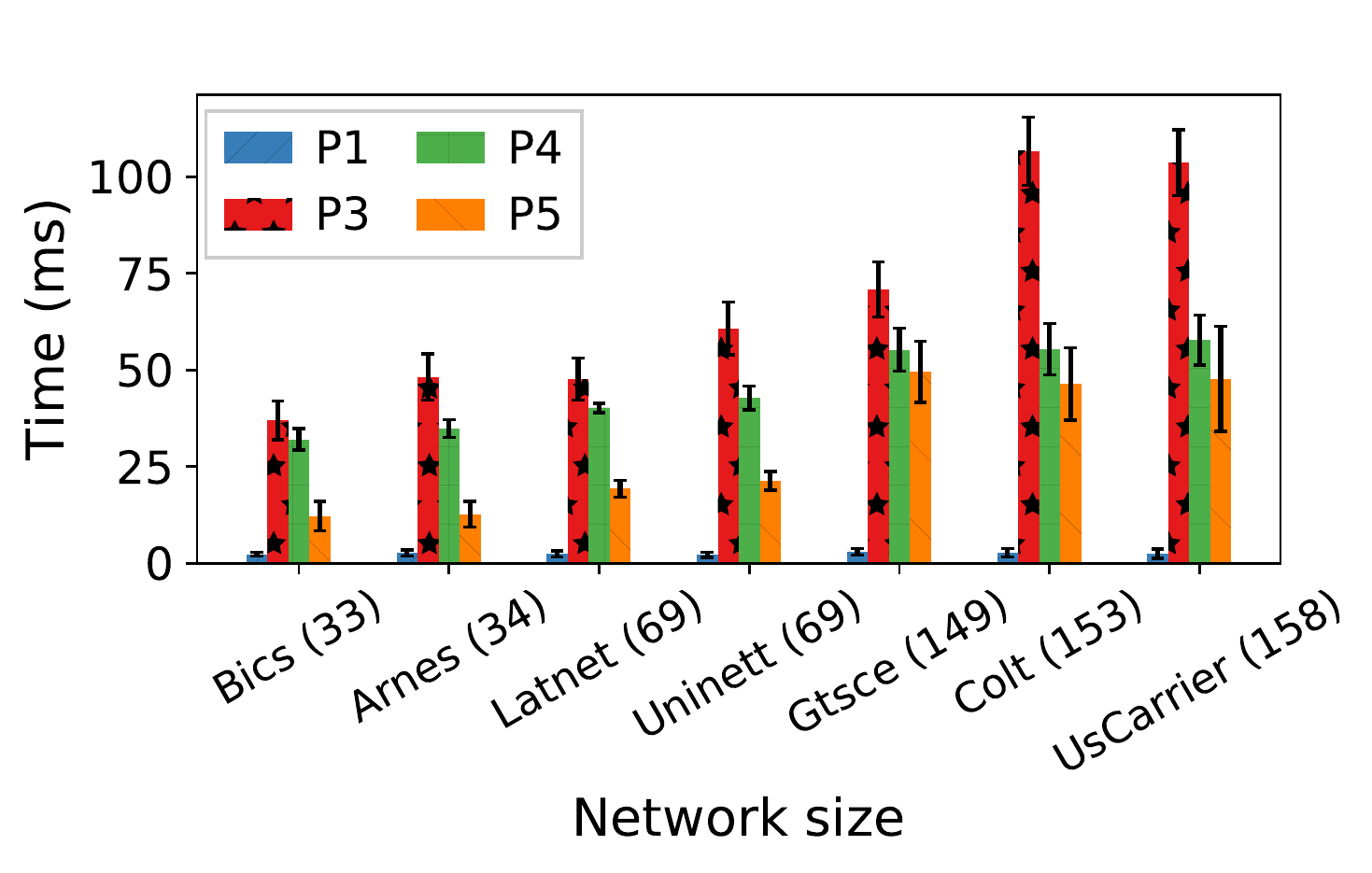}
	\compactcaption{Performance on scale (topology zoo networks)}
	\label{fig:topologies}
\end{figure}

Our implementation of \Name is written in Java. We use Batfish~\cite{batfish} to parse router configurations. From these, we generate our multilayer graphs (\secref{sec:multigraph}). We implemented all our verification algorithms (\secsref{sec:pathset}{sec:path}) in Java. \Name uses Gurobi~\cite{gurobi} to solve our ILPs. In all, this amounted to $\approx7$K lines of code. We evaluate \Name on a variety of issues:
\begin{compactitemize}
\item How quickly can \Name verify different policies?
\item How does \Name perform relative to
  Minesweeper~\cite{minesweeper}?
\item How does \Name's performance scale with network size?
\end{compactitemize}
All our experiments were performed on machines with 40 core 2.2 GHz Intel Xeon Silver Processors and 192 GB RAM.

\subsection{Network Characteristics}\label{subsec:nw}
In our evaluation, we use configurations from (a) 4 university
networks, (b) 34 real datacenter networks operated by a large OSP,
and, (c) 7 networks from topology zoo dataset~\cite{zoo}. The
university networks have 9 to 35 devices.  The university networks are the richest in terms
of configurations constructs. They support eBGP, iBGP, OSPF, Static routes, ACLs/Filters, community, local preference, 
MPLS+VRFs and VLANs. The datacenter networks have between 2 and 24
devices. They do not employ local preference, MPLS or VLANs. Finally,
the topology zoo networks have between 33 and 158 devices. The configs
for these networks were synthetically generated using
NetComplete~\cite{netcomplete}. These generated configs do not have
Static routes, ACLs, MPLS or VLANs as Netcomplete cannot model them. \tabref{tab:conf} shows what percentage of networks in these datasets support each network protocol/modifier.

\figref{fig:graph} characterizes the size of all the multilayer graphs
generated by \Name for these networks. It first shows the number of
nodes and edges used to represent the {\emph base} multigraph of these
networks. We observe two outliers in both~\figref{fig:graph}a and
\figref{fig:graph}b. These occur for networks $Uni-2$ (24 devices) and
$Uni-3$ (26 devices), from the university dataset. These networks have
multiple VRFs and VLANs, and \Name creates nodes (and edges between
these nodes) per VRF/VLAN per routing process. Note also that for the
other networks, the number of routing processes per device
varies. Hence, the number of nodes and edges do not monotonically
increase with network size.

\setlength{\textfloatsep}{0.35cm}
\setlength{\floatsep}{0.35cm}

\textbf{Policies.}
We consider five types of polices: (\policyBlock) always
unreachable, (\policyWay) always waypointing, (\policyFail) always reachable
with $< K$ failures, (\policyBound) always bounded length, and (\policyPref) path preference. Recall from other sections: \policyBlock and \policyWay use \tdfs; \policyPref uses \pvp and \tyen algorithms; \policyFail uses ILP and \tyen.; And \policyBound uses another ILP. Using these policies, we evaluate the performance of all our algorithms. 

\subsection{Verification Efficiency}\label{subsec:ver}
We examine how efficiently \Name can construct and
verify these multilayer graphs. First, we evaluate the time required
to generate these graphs. We use configurations from all the
networks. \figref{fig:gen} shows the time taken to generate the {\em
  base graphs} and per {\em traffic class-specific graph} for all
networks. \Name can generate these graphs, even for large networks, in
$\approx$ 30 $\mu$s. The time to generate the traffic-class graph from
the base graph is atmost 3 $\mu$s on average per traffic-class.

\eat{
\begin{figure}
	\includegraphics[scale=0.35]{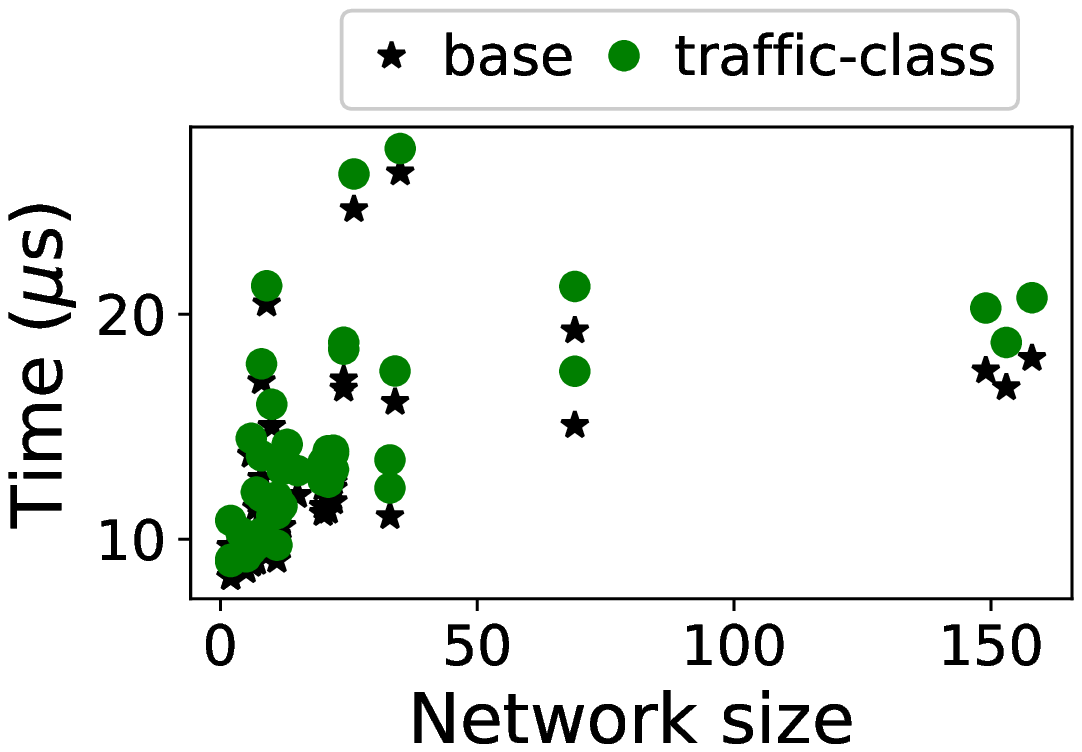}
	\compactcaption{Multilayer graph generation time of (all networks)}
	\label{fig:gen}
\end{figure}

\begin{figure}
	\includegraphics[scale=0.35]{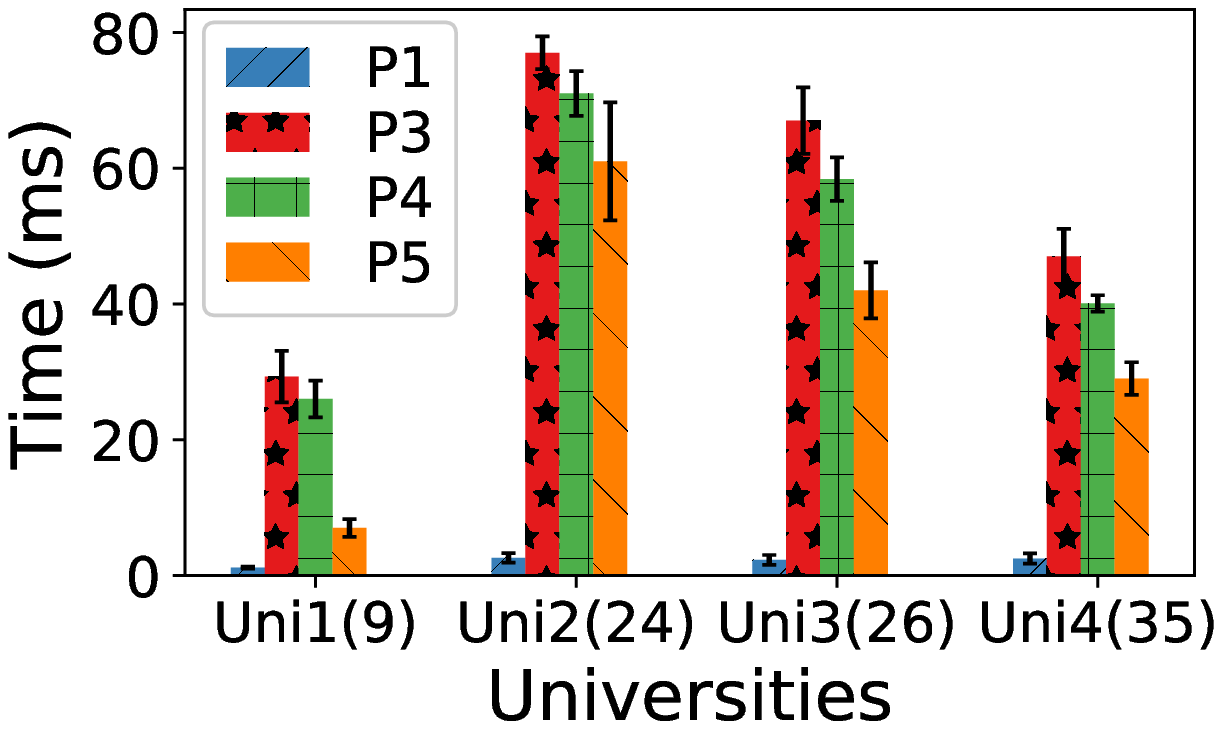}
	\compactcaption{Verify policies on university configs}
	\label{fig:univs}
\end{figure}
}

Next, we examine how efficiently \Name can verify various
policies. Since the university networks are the richest in terms of
configuration constructs, we use them in this
experiment. \figref{fig:univs} shows the time taken to verify policies
\policyBlock, \policyFail, \policyBound, and \policyPref. In this
and all the remaining experiments, the values shown are the median
taken over {\em 100 runs} for {\em 100 different traffic
  classes}. Error bars represent the std. deviation.

We observe that \policyBlock can be verified in less than 3 ms. Since
it uses a simple polynomial-time graph traversal algorithm (\tdfs), it
is the fastest to verify among all policies. The time taken to verify
\policyPref is higher than \policyBlock, because \pvp and \tyen
algorithms are more complex, as they run our path vector protocol to
convergence to find paths (and in \tyen's case the protocol is invoked
many times). Finally, \policyFail and \policyBound, both use an ILP
and, as expected, are slowest to verify. However, they can still be
verified in $\approx$ 80 ms per traffic class.

Although $Uni2$ and $Uni3$ have fewer devices than $Uni4$, they
have more nodes and edges in their \Name graphs
(\secref{subsec:nw}). Hence it takes longer to verify policies on them.

\subsection{Comparison with Minesweeper}\label{subsec:minesweeper}
Next, to put our performance results in perspective, we compare \Name
with Minesweeper. In this experiment we use the real datacenter
networks. We consider policies \policyBlock, \policyWay, \policyBound,
and \policyPref. In Minesweeper, we have to specify the number of
failures $K$; Minesweeper then verifies if the policy holds as long as
there are $\leq K$ failures. To verify a property under all failure
scenarios, we set the value of $K$ to {\em number of physical links in
  the network - 1}. \figref{fig:minesweeper} (a, b, c, and d) shows
the time taken by \Name and Minesweeper to verify these
policies. \figref{fig:failspeed} (a, b, c, and d) shows the speedup
provided by \Name for each of these policies.

For policies that use \tdfs (\policyBlock and \policyWay), \Name's speedup is as high as $600$-$800X$. For \policyBound, the speedup is as high as $50X$.

\policyPref is the only policy where speedup does not increases with
network size. This is because larger networks have longer path
lengths and more possible candidate paths. Both of these affect the
complexity of the \tyen algorithm. The number of times \tyen invokes \pvp 
increases significantly with network size. Hence the
speedup for \policyPref is relatively less, especially at larger network sizes.

Note that since the number of nodes and edges did not monotonically
increase with network size (\secref{subsec:nw}), the time to verify
these policies does not monotonically increase either.

Next, we compare the performance of \Name and Mines- weeper for the same
policies but without failures, e.g. ``currently reachable'' instead of
``always reachable''. \Name verifies these policies by generating the
actual path using \pvp. \figref{fig:nofail} (a, b, c, and d) shows the
speedup provided by \Name for each of these policies. Even for no
failures, \Name significantly outperforms Minesweeper across all
policies. Minesweeper has to invoke the SMT solver to find a
satisfying solution even in this simple case.

To shed further light on \Name's benefits w.r.t. Mineswe- eper, we
compare the number of variables used by Mineswee- per's SMT encoding and
\Name's ILP encoding to verify \policyFail and \policyBound. In this
experiment, we track {\emph reduction ratio}, which is the number of
variables in Minesweeper divided by number of variables in \Name. This
is shown in \figref{fig:vars}. As expected, \Name uses significantly
fewer variables. Also, \policyFail uses more variables than
\policyBound. Hence it has a lower reduction ratio. \Name uses integer
variables only for the aspects that matter towards the property in
question, where Minesweeper uses binary and integer variables
throughout its general encoding, irrespective of the property in
question. This is one reason for Minesweeper's poor performance.

\subsection{Scalability}\label{subsec:scale}

Until now, all our experiments were on small networks. In this
section, we evaluate \Name's performance on large networks from the
topology zoo. \figref{fig:topologies} shows the time taken to verify
policies \policyBlock, \policyPref, \policyFail, and,
\policyBound. \Name can verify these policies in $<$ 0.12 s, even in
these large networks.

As expected, verifying \policyBlock (\tdfs) is significantly faster than all other policies, and it is very low across all network sizes. However, for larger networks, time to verify \policyPref (\tyen) is as high as \policyBound. Again, this due to larger networks having longer and more candidate paths. Large networks also have high diversity in terms of path lengths. Hence, we see more variance in the time to verify \policyPref compared to other policies.

For large networks, the time to verify \policyFail is significantly
higher than other policies. This happens because \policyFail's ILP
formulation becomes more complex, in terms of number of variables, for
such large networks.

\section{Related Work}\label{related}

We surveyed various related studies in detail, in earlier
sections. Here, we survey others that were not covered earlier.

Aside from Minesweeper and ARC, there are other control plane
verification tools that attempt to verify policies against various
environments (failures or advertisements). ERA~\cite{era} symbolically
represents control plane advertisements which it propagates through a
network and transforms it based on how routers are
configured. ERA is useful to verify reachability against arbitrary
external advertisements, but it does not have the full coverage of
control plane constructs as \Name or Minesweeper to analyze a range of
policies. Bagpipe~\cite{bagpipe} is similar in spirit to Minesweeper
and \Name, but it only applies to a network that only runs
BGP. FSR~\cite{fsr} focuses on encoding BGP path preferences.

Batfish~\cite{batfish} and C-BGP~\cite{cbgp} are control plane
simulators. They analyze the control plane's path computation as a
function of a given environment, e.g., a given failure or an incoming
advertisement, by conducting low level message exchanges, emulating
convergence, and creating a concrete data plane. \Name also conducts
simulations of the control plane; but, for certain policies, \Name
can explore multiple paths at once via graph traversal and avoid
protocol simulation. For other policies, \Name only simulates a
simplified protocol (SPVP) running over a multi-layer multi-attributed
graph, and the simulations are conducted in parallel per traffic class.

\section{Limitations} 
One limitation of graph-based control plane
models is that they cannot symbolically model advertisements, which is
easier to do for SMT-based tools. \Name shares this drawback. This
means that \Name cannot exhaustively explore if there exists an
external advertisement that could potential lead to a property
violation; \Name can only exhaustively explore link failures.  \Name
would have to be provided a concrete instantiation of an advertisement;
in such a case, \Name can analyze the network under the given
advertisement and determine if any policies can be violated.

A related issue is that \Name cannot be applied to verify control
plane equivalence: Two control planes are equivalent, if the behavior
of the control planes (paths computed) is the same under all
advertisements and all failure scenarios.

In essence, while \Name can replace Minesweeper for a vast number of
policies, it is not a universal replacement. Minesweeper's
SMT-encoding is useful  to explore advertisements.
\section{Conclusion}

While existing graph-based control plane abstractions are fast, they
are not as general. SMT-based abstractions are general, but not
fast. In this paper, we showed that graphs can be used as the basis
for general and fast network verification. Our insight is that, rich,
multi-layered graphs, coupled with algorithmic choices that are
customized per policy can help achieve the best of both worlds. Our
evaluation of a prototype (which we will release open source) shows
that we can offer 10-600X better speed than state-of-the-art, scale
gracefully with network size, and model all key features found in
network configurations in the wild. 
{\em This work does not raise any ethical issues.}

{
	\raggedright
	\balance
	\bibliography{main}
	\bibliographystyle{abbrv}
}
\appendix
\section{Other Policies}\label{sec:otherPols}
Some of the other policies that \Name can verify are listed below:

\textbf{Always Chain of Waypoints (\policyChain).}  Similar to waypointing, we remove
nodes associated with each waypoint, one at a time. Using \tdfs, we
check if nodes associated with one of the preceding waypoints in the
chain can reach nodes associated with one of the following waypoints
in the chain.

\textbf{Equal Bound (\policyEqual).}  This policy checks that all paths from $src$ to
$dst$ are of the same length. The objective of the ILP
in~\secref{subsec:bound} can be changed to find the shortest path
length. If the longest and shortest path length varies, then this
policy is violated.

\textbf{Always isolated (\policyIso).}  Two traffic classes are always isolated if
they never traverse the same link. ARC~\cite{arc} verified this policy
by checking if two \emph{traffic class-specific} graphs have any edge
in common or not. Similarly, in \Name, we check if the two
\emph{traffic class-specific} graphs have any hedge in common.

\textbf{Multipath Consistency (\policyMul).}  Multipath consistency is violated
when traffic is dropped along one path but blocked by an ACL on
another. To support multipaths in \pvp, we change the \bestgen variable
to track multiple best advertisements. Using \pvp, \Name can identify
the number of best paths. We run \pvp on graphs with and without
removing edges for ACLs. If the number of paths varies with and
without ACLs, then the policy is violated.

\textbf{Always no black holes (\policyBlack).}  Black holes occur when traffic gets
forwarded to a router that does not have a valid forwarding
entry. Blackholes are caused by i) ACLs: router advertises routes for
traffic that are blocked by their ACLs; ii) static routes: the
next-hop router of a static route cannot reach $dst$. \Name uses \tdfs
to check these conditions. For (i) \Name first creates the graph
without removing edges for ACLs. Let $R$ be the router with a blocking
ACL. If $src$ can reach router $R$ and $R$ can reach $dst$ (using
\tdfs), then traffic will reach router $R$ under some failure, and
then get dropped because of the ACL. For (ii) if $src$ can reach the
router with static route and the next-hop router \emph{cannot reach}
$dst$, then the traffic gets dropped.
\section{Proofs}\label{sec:proofs}
\subsection{Prohibited Paths}\label{subsec:proh}
Any path with $>$ 2 consecutive untainted nodes is a prohibited path

\setcounter{theorem}{0}

\begin{theorem}\label{thr:proh}
The approach specified in~\secref{subsec:prohibit} correctly identifies all prohibited path.
\end{theorem}

\textbf{Proof:} We will prove this by contradiction.


In a router, each routing process will have its own Routing Information Base (RIB) that has routes learned by that routing process. The routes of all RIBs are also inserted in the Forwarding Information Base (FIB) of that router. Hence, the ``f'' edge exists between all routing process node and the fib node for each router.

The progression of taints is modelled based on $G_{RIB}$ graph of~\cite{xie2005static}. In $G_{RIB}$, edge adjacency exists only for RIB adjacency. Similarly taints are spread only across RIB adjacent processes. Similar to~\cite{xie2005static}, taints cross intra-device routing processes only with redistribution.

A prohibited path can exist only if you traverse a router that does not have a valid fib entry. Assume such a path exists and is not identified by Tiramisu's tainting strategy. Such paths have to satisfy the following three criterias: (i) there has to be a router where path goes from tainted node to untainted node, i.e. from node $W$ representing routing process or FIB (e.g. $C_{bgp}$, $C_{fib}$ in~\figref{fig:correlated}) that has forwarding entry, to node $X$ representing routing process that does not have forwarding entry ($C_{ospf}$); (ii) node $X$ has to be followed by a node $Y$ ($D_{ospf}$) representing another router with the same routing process; and (iii) node $Y$ has to be succeeded by node $Z$ which will either be a node representing node $Y$'s router's FIB ($D_{fib}$) or another router with the same routing process as Y ($A_{ospf}$). Here both these nodes have to be untainted. However, this ends up with a path with three untainted nodes, which are classified by Tiramisu as prohibited. This contradicts the assumption that this path is not identified by Tiramisu.

\subsection{TPVP}\label{subsec:tpvp_proof}

\begin{theorem}\label{thr:tpvp}
Assuming the network control plane converges, \pvp always finds the exact path taken in the network under any given failure scenario.
\end{theorem}

\textbf{Proof:} We leverage correctness of Minesweeper and SPVP to establish the correctness of \pvp. In SPVP, there is no restriction on order in which messages are processed by different routers. As long as there exists even one stable path, SPVP will find it irrespective of the order of message processing. In \pvp (line 5), we process messages in a fixed round-robin order. Since any ordering of messages in SPVP leads to a valid solution, a fixed ordering of messages should also lead to a valid solution. 

The body of the loop code of \pvp is equivalent to SPVP. The updateCost function is a function that was not mentioned in the original SPVP algorithm. In the original SPVP algorithm, rib-in (u $\Leftarrow$ v) represented paths received from neighbor v. And the properties of those paths e.g. local-pref, ospf-weight etc were assigned based on import and export policies. Minesweeper also modelled rib-in (u $\Leftarrow$ v) as import and export policy constraints. Tiramisu's updateCost function also updates path properties based on import and export policy constraints, similar to Minesweeper~\cite{minesweeper} and SPVP~\cite{griffin2002stable}. 

\subsection{TDFS}\label{subsec:tdfs_proof}
 
\begin{theorem}\label{thr:tdfs}
TDFS traverses all real network paths that could materialize under some failure scenario, and does not traverse any path that cannot materialize under any failure scenario.
\end{theorem}

\textbf{Proof:} Let path $P$ be the path traversed by a particular src-dst pair in the actual network under some failure. Now assume this path is not traversed by \tdfs because it does not exist in the graph. This contradicts \thrref{thr:tpvp}, which showed that \pvp always finds the exact path used in the actual network, which implies the path must exist in the graph. Thus, \tdfs will traverse all network paths that could materialize under some failure scenario.

Now assume there exists some path $P'$ in the actual network that also exists in the graph. Also assume it is not traversed by \tdfs. This means this path $P'$ contains three or more consecutive tainted nodes. This contradicts \thrref{thr:proh} as it cannot be a real network path. Thus, \tdfs does not traverse any path that cannot materialize under any failure scenario.
\section{\Name Min-cut}\label{subsec:min-cut-appendix}

\tabref{tab:function} lists the boolean indicator variables and
functions used in encoding of the ILP for the property in
\secref{subsec:min-cut}. We now provide the full ILP with detailed constraints;
we repeat the description of the constraints to ensure ease of
reading. Recall, edges in our graphs represented the flow of traffic
from $src$ to $dst$. For ease of understanding, in specifying the ILP,
we reverse the edges to represent flow of advertisement from $dst$ to
$src$.


\begin{table}[!htb]
\begin{centering}
  \small
  \setlength\tabcolsep{3pt} 
\begin{tabular}{p{0.2cm}|p{2.0cm}|p{5.8cm}}
\hline
& \textbf{Name} & \textbf{Description}\\ \hline
\parbox[t]{2mm}{\multirow{6}{*}{\rotatebox[origin=c]{90}{Variable}}} & $F_e$ & set as 1 if edge $e$ fails\\
& $A_e$ & set as 1 if advertisement propagates on edge $e$\\
& $R_n$ & set as 1 if advertisement reaches node $n$\\
& $B_e$ & set as 1 if edge $e$ carries blocked community\\
& $P_e$ & set as 1 if edge $e$ only propagates advertisement that create prohibited paths\\
& $C_{n,c}$ & set as 1 if node $n$ forwards community $c$\\
\hline
\parbox[t]{2mm}{\multirow{13}{*}{\rotatebox[origin=c]{90}{Function}}} & $nodes$ & returns all nodes of graph\\ 
& $edges$             & returns all edges of graph\\ 
& $dEdges$             & returns all inter-device edges of graph\\ 
& $phyEdges$             & returns all physical edges of graphs\\ 
& $uEdges$             & returns all edges of graph with untaint start and end nodes\\ 
& $oNodes$             & returns all nodes except $src$ and $dst$\\

& $iE(n)$             & returns incoming edges of node $n$\\ 
& $oE(n)$             & returns outgoing edges of node $n$\\ 
& $iUE(n)$             & returns incoming edges from untainted neighbors of node $n$\\ 
& $iN(n)$             & returns start nodes of all incoming edges of node $n$\\ 
& $ac(n)$             & returns communities added by node $n$\\ 
& $rc(n)$             & returns communities removed by node $n$\\ 
& $bc(n)$             & returns communities blocked on node $n$\\ 
& $oc(n)$             & returns communities $\notin$ [$ac(n), bc(n), rc(n)$] \\ 
& $start(e)$             & returns start node of edge $e$ \\ 
& $end(e)$             & returns end node of edge $e$ \\ 
\hline
\end{tabular}
\end{centering}
\caption{\textit{Variables and Functions}}
\label{tab:function}
\end{table}

\textbf{ILP}
The objective of the ILP is to minimize the number of physical link failures required to disconnect the $src$ from $dst$. All edges belonging to hedge $i$ will share the same $F_i$ variable.

\begin{equation} 
\textbf{Objective:} \tab  \texttt{minimize} \sum_{i \in  phyEdges} F_i
\end{equation}

\textbf{Advertisement Constraints} 
We first discuss the constraints added to represent reachability of advertisements. The base constraints state that the $dst$ originates the advertisement. To disconnect $dst$ from $src$, the advertisement must not reach the $src$.

\begin{equation}\label{eq:dst}
R_{dst} = 1
\end{equation}
\vspace*{-1em}
\begin{equation}\label{eq:src}
R_{src} = 0
\end{equation}

For other nodes, we add the constraint that advertisement reaches $n$ if it gets propagated ($A_e$) on any of its incoming edge $e$. \begin{multline} \label{eq:reach_cons}
\forall n \in oNodes, R_n = \bigvee_{ e \in iE(n)} A_e
\end{multline}

Note that logical AND, OR operators can be represented as constraints in ILP.

Next, we discuss constraints on propagation of advertisements. An advertisement can propagate through an edge $e$, if it reaches the start node ($n$) of the edge ($R_n$), the physical edge does not fail ($\neg F_e$), the advertisement does not carry a community ($\neg B_e$) that is blocked on this node, and if edge $e$ carries atleast one advertisement that does not create a prohibited path ($\neg P_e$). This is represented as
\vspace*{-0.5em}
\begin{multline} \label{eq:flow_cons}
\forall n \in oNodes, \forall e \in iE(n): \\ A_e = R_n \land \neg F_e \land \neg B_e \land \neg P_e
\end{multline}

\textbf{Community Constraints} 
The base constraints state that each nodes that adds the community forwards that community and each nodes that removes that community does not forward it.
\begin{multline} \label{eq:add_comm}
\forall n \in nodes, \forall c \in ac(n): \\ C_{n,c} = 1
\end{multline}
\begin{multline} \label{eq:rem_comm}
\forall n \in nodes, \forall c \in rc(n): \\ C_{n,c} = 0
\end{multline}

For other nodes, we add the constraint that node $n$ forwards a community $c$ iff any of its inbound neighbors ($iN(n)$) forwards that community to node $n$.
\begin{multline} \label{eq:oth_comm}
\forall n \in nodes, \forall c \in oc(n): \\ C_{n,c} = \bigvee_{ i \in iN(n)} C_{i,c}
\end{multline}

Finally, we add the constraint that an edge $e$ carries a blocked community iff the start node of edge $e$ forwards any community that is blocked by end node of edge $e$.
\begin{multline} \label{eq:block_comm}
\forall e \in edges: \\ B_e = \bigvee_{c \in bc(end(e))} C_{start(e), c}
\end{multline}

\textbf{Prohibited Path Constraints} 
An edge $e$, that only propagates advertisements that create prohibited paths, is an edge that satisfies the condition that (\eqnref{eq:proh}) the start and end nodes of edge $e$ are untainted ($uEdges$). Furthermore the start node receives advertisement only on edges from untainted neighbors ($A_{ue}$) and not on edges from any other neighbor ($A_{ne}$). Such edges always create subpaths with three consecutive untainted nodes.
\begin{multline} \label{eq:proh}
\forall e \in uEdges: P_e = \left( \bigvee_{ue \in iuE(start(e))} A_{ue} \right) \land \\ \neg  \left( \bigvee_{ne \in \neg iuE(start(e))} A_{ne} \right)
\end{multline}




\section{\Name Longest Path}\label{subsec:bound-appendix}

We now specify the ILP corresponding to \secref{subsec:bound} in detail. For
ease of reading, we also provide the full description of the objective
and the constraints.

Recall that to verify \policyBound, we propose another ILP whose
objective is to maximize the number of inter device edges ($dEdges$)
traversed by an advertisement ($A_i$). Note again that, the path
traversed by the advertisement is the opposite of traffic flow.  Our
objective is as follows:

\begin{equation} 
\textbf{Objective:} \tab  \texttt{maximize} \sum_{i \in dEdges} A_i
\end{equation}

\textbf{Single Path Constraints} To ensure that only one path
gets advertised, and that the $dst$ sends the advertisement and the
$src$ receives the advertisement, we add the  constraints in

\begin{equation}\label{eq:dst2}
\sum_{ out \in outEdge(dst)} A_{out} = 1
\end{equation}

\begin{equation}\label{eq:src2}
\sum_{ in \in inEdge(src)} A_{in} =  1
\end{equation}

For other nodes, we add the flow conservation property, i.e. sum of incoming flows is equal to outgoing flows

\begin{multline}\label{eq:flowconser}
\forall n \in oNodes: \sum_{ in \in iE(n)} A_{in} =  \sum_{ out \in oE(n)} A_{out}
\end{multline}

\textbf{Advertisement Constraints} Next, we add constraints on
propagation of advertisements. An advertisement can be bloc- ked on edge
$e$ if it satisfies the community ($B_e$) and path
prohibition ($P_e$) constraints. These are similar to
\eqnref{eq:block_comm} and \eqnref{eq:proh}.

\begin{multline}\label{eq:extracons}
\forall e \in edges: A_e \leq  \neg (B_e \lor P_e)
\end{multline}

\end{document}